\pdfoutput=1

\documentclass[%
reprint,
superscriptaddress,
nofootinbib,%so that there is no "Note" in the footnote
amsmath,amssymb,amsfonts
aps,
pre,
letterpaper,
]{revtex4-1}

\usepackage{tabularx}
\usepackage{amsopn}
\usepackage{bm}% bold math
\usepackage{booktabs}
\usepackage{epsfig}
\usepackage{subfigure}
\usepackage{amsfonts}
\usepackage{comment}%for comments
\usepackage{enumerate}
\usepackage{mathrsfs} % for huati letter
\usepackage[perpage,symbol]{footmisc}   \setfnsymbol{wiley}   %注释不使用1,2,3，而使用*,**等
\usepackage{multirow} %用于table中有的单元格合并两行
\usepackage{booktabs}% http://ctan.org/pkg/booktabs  %用于table内的itemize
\newcommand{\tabitem}{~~\llap{\textbullet}~~}  %用于table内的itemize
\DeclareMathOperator*{\argmax}{arg\,max}

\begin{document}
\title{Detecting network communities beyond assortativity-related attributes}
\author{Xin Liu}
\email[To whom correspondence should be addressed.\\E-mail: ]{tsinllew@ai.cs.titech.ac.jp}
%\email{E-mail: tsinllew@ai.cs.titech.ac.jp}
\affiliation{Tokyo Institute of Technology, 2-12-1 Ookayama, Meguro, Tokyo, 152-8552 Japan}
\affiliation{CREST JST, K's Gobancho, 7, Gobancho, Chiyoda, Tokyo, 102-0076 Japan}
\affiliation{Wuhan University of Technology, 122 Luoshi Road, Wuhan, Hubei, 430070 China}
%\email[Corresponding author]{tsinllew@ai.cs.titech.ac.jp}
%\thanks{Corresponding author}
\author{Tsuyoshi Murata}
\affiliation{Tokyo Institute of Technology, 2-12-1 Ookayama, Meguro, Tokyo, 152-8552 Japan}
\author{Ken Wakita}
\affiliation{Tokyo Institute of Technology, 2-12-1 Ookayama, Meguro, Tokyo, 152-8552 Japan}
\affiliation{CREST JST, K's Gobancho, 7, Gobancho, Chiyoda, Tokyo, 102-0076 Japan}

\begin{abstract}
In network science, assortativity refers to the tendency of links to exist between nodes with similar attributes. In social networks, for example, links tend to exist between individuals of similar age, nationality, location, race, income, educational level, religious belief, and language. Thus, various attributes jointly affect the network topology. An interesting problem is to detect community structure beyond some specific assortativity-related attributes $\rho$, i.e., to take out the effect of $\rho$ on network topology and reveal the hidden community structure which are due to other attributes. An approach to this problem is to redefine the null model of the modularity measure, so as to simulate the effect of $\rho$ on network topology. However, a challenge is that we do not know to what extent the network topology is affected by $\rho$ and by other attributes. In this paper, we propose Dist-Modularity which allows us to freely choose any suitable function to simulate the effect of $\rho$. Such freedom can help us probe the effect of $\rho$ and detect the hidden communities which are due to other attributes. We test the effectiveness of Dist-Modularity on synthetic benchmarks and two real-world networks.
\end{abstract}

\maketitle

\section{Introduction}
Many social, biological, and information systems can be described by networks, where nodes represent fundamental entities of a system, such as individuals, users, genes, web pages, and links represent relations or interactions between the entities \cite{BarabasiNetSci}. In recent years, there has been a surge of interest in the analysis of networks. A highly discussed topic is community detection --- the detection of groups of network nodes, known as communities, within which links are dense, but between which links are sparse \cite{GirvanNewmanCommuityDefinition}. Community detection \cite{FortunatoCommunityDetectionReview,NewmanCommunityReview,DanonCompareCommunityAlgo,LancichinettiCommunityAlgoAnalysis,OrmanComparativeEvaluationCommunityDetectionAlg} is considered as a crucial step towards inferring function units of the underlying system, such as collections of pages on closely related topics on the web or groups of people with common interest in social media.

People observed that in real-world networks links tend to exist between nodes with similar attributes. For example, in social networks individuals commonly choose to associate with others of similar age, nationality, location, race, income, educational level, religious belief, and language as themselves. This tendency is known as \textit{assortativity} (also known as assortative mixing or homophily) \cite{NewmanAssortativeMixing,NewmanMixingPatterns,McPhersonHomophily,PapadopoulosPopularitySimilarityInGrowingNetwork}. To see whether an attribute is assortativity-related, or whether it is correlated with the network topology, Bavaud proposed a modes permutation test \cite{BavaudTestingSpatialAutocorrelation}.

Assortativity indicates that various attributes jointly affect the network topology, either directly or indirectly. An interesting problem is to detect community structure beyond some specific attributes, represented by a variable vector $\rho$. In other words, the goal is to take out the effect of $\rho$ on network topology and reveal the hidden community structure which are due to other attributes, represented by a variable vector $\bar{\rho}$. Note that $\rho$ is observable, while $\bar{\rho}$ may contain some latent variables. For example, one may be interested in studying the community structure which is not due to age, but due to religious belief, educational level, income, and some attributes which are not observed in social networks. A challenge of this problem is that we do not know to what extent the network topology is affected by $\rho$ and $\bar{\rho}$. Moreover, some attributes in $\rho$ and $\bar{\rho}$ can be correlated. This makes the problem even more complex, since it becomes more difficult to disentangle $\rho$ and $\bar{\rho}$ \cite{CerinaSpatialCorrelation}.

The popular community detection method which relies on optimization of a quantity measure called NG-Modularity (Newman-Girvan modularity) \cite{NewmanModularityDefinition} cannot solve the above problem. This is because the definition of NG-Modularity does not take node attributes $\rho$ into account. The definition of NG-Modularity involves a comparison between the observed network and a \textit{null model}. In order to solve the problem, this null model should be redefined to simulate the effect of $\rho$ on network topology, so that such effect can be taken out when compared to the observed network. Following this idea, Expert et al. defined a new null model based on an empirically determined probability distribution and proposed Spa-Modularity \cite{ExpertSpatialCommunity}. However, recent experiments by Cerina et al. showed that Spa-Modularity still cannot handle this problem well, especially when there is a correlation between $\rho$ and $\bar{\rho}$ \cite{CerinaSpatialCorrelation}.

In this paper, we extend NG-Modularity and propose Dist-Modularity. In particular, we define a general null model which allows us to freely choose any suitable function to simulate the effect of $\rho$ on network topology. Such freedom can help us probe the effect of $\rho$ and detect the hidden communities which are due to $\bar{\rho}$. We use synthetic networks and two real-world networks to demonstrate the effectiveness of Dist-Modularity. We analyze the reasons for why Spa-Modularity fails in these examples.

The rest of the paper is organized as follows. Section~\ref{sec2} reviews NG-Modularity and Spa-Modularity. Section~\ref{sec3} introduces our new null model and Dist-Modularity. Section~\ref{sec4} presents experiments, followed by a conclusion in Section~\ref{sec5}.

\section{Related Work}\label{sec2}
In this section, we give a review of NG-Modularity and Spa-Modularity. Before this, let us first introduce the prototype of these two measures --- the Modularity \cite{NewmanModularityDefinition,NewmanModularityEigenvectors}.

\subsection{Modularity}\label{sec2.1}
Modularity is a quantity measure for evaluating the quality of a partition of a network into communities. The definition of Modularity compares the fraction of within-community links in the observed network minus the expected value of that fraction in some equivalent randomized network. This randomized network is called the null model, which serves as a reference. The mathematical expression of Modularity in an undirected network reads
\begin{eqnarray}
\mathrm{Q}(\mathcal{L})=\frac{1}{2m} \sum_{i,j=1}^n (A_{ij}-P_{ij})\,\delta(l_{i},l_{j}), \label{eq1}
\end{eqnarray}
where $n$ is the number of nodes, $m$ is the number of links, $\mathcal{L}=\{l_1,l_2,\cdots,l_n\}$ is a partition, with element $l_{i}$ indicating the community membership of the $i$-th node $v_i$, ${A}_{ij}$ is the number of links between $v_i$ and $v_j$ in the observed network, ${P}_{ij}$ is the expected value of that number in the null model, and $\delta$ is the Kronecker's delta.

To make Eq.~\eqref{eq1} significant, the number of links in the null model should equal that number in the observed network. That is,
\begin{eqnarray}
\sum_{i,j=1}^n P_{ij} = \sum_{i,j=1}^n A_{ij} = 2m. \label{eq2}
\end{eqnarray}
Apart from the constraint \eqref{eq2}, there is some freedom about choosing the null model, and different null models produce variants of Modularity. According to Eq.~\eqref{eq1} and \eqref{eq2}, it is clear that $\mathrm{Q} \in [-1,1]$. For a given network, the higher the $\mathrm{Q}$, the better the partition $\mathcal{L}$.

\subsection{NG-Modularity}\label{sec2.2}
A popular choice of the null model proposed by Newman and Girvan \cite{NewmanModularityDefinition,NewmanModularityEigenvectors} is the configuration model \cite{NewmanRandomGraphArbitraryDegree}, which preserves the degree sequence of the observed network. Specifically, the expected number of links between $v_i$ and $v_j$ in this null model is
\begin{eqnarray}
P_{ij}^{\mathrm{NG}}=k_ik_j/2m, \label{eq3}
\end{eqnarray}
where $k_i=\sum_{j=1}^n A_{ij}$ is the degree of $v_i$. Replacing $P_{ij}$ in Eq.~\eqref{eq1} by $P_{ij}^{\mathrm{NG}}$ gives NG-Modularity ($\mathrm{Q}^{\mathrm{NG}}$), which is widely used in practice.

\subsection{Spa-Modularity}\label{sec2.3}
The configuration model used in NG-Modularity does not take attributes into account. To detect community structure beyond $\rho$, the null model should be redefined to simulate the effect of $\rho$ on network topology, so that such effect can be taken out when compared to the observed network. For this reason, Expert et al. defined a new null model (originally for simulating the effect of space on network topology) \cite{ExpertSpatialCommunity}. Specifically, the expected number of links between $v_i$ and $v_j$ in this null model is
\begin{eqnarray}
P_{ij}^{\mathrm{Spa}}=h_ih_jp(d_{ij}), \label{eq4}
\end{eqnarray}
where $d_{ij}$ denotes the distance between $v_i$ and $v_j$ in terms of $\rho$, $h_i$ indicates the importance of $v_i$, $p(d)$ is the probability that two nodes are connected at a distance $d$, and can be obtained empirically from the observed network by
\begin{eqnarray}
p(d) = \frac{ \sum_{i,j|d_{ij}=d} A_{ij} } { \sum_{i,j|d_{ij}=d} h_ih_j }. \label{eq5}
\end{eqnarray}
Replacing $P_{ij}$ in Eq.~\eqref{eq1} by $P_{ij}^{\mathrm{Spa}}$ gives Spa-Modularity ($\mathrm{Q}^{\mathrm{Spa}}$).

From Eq.~\eqref{eq4} and \eqref{eq5}, we can derive that
\begin{eqnarray}
\sum_{i,j|d_{ij}=d} P_{ij}^{\mathrm{Spa}} = \sum_{i,j|d_{ij}=d} A_{ij}. \label{eq6}
\end{eqnarray}
This indicates that the number of links between nodes at distance $d$ in the Spa-Modularity null model is the same as that number in the observed network. In other words, this null model assumes that only $\rho$ affects network topology, and thus it simulates the effect of $\rho$ as what can be observed in network topology.

\section{Dist-Modularity}\label{sec3}
In this section, we first propose a new null model which allows us to freely choose any suitable function to simulate the effect of $\rho$ on network topology, and then present Dist-Modularity for detecting communities beyond $\rho$. For simplicity, we only consider the case of undirected networks.

\subsection{Our New Null Model}\label{sec3.1}
We propose the following new null model. In this model, the expected number of links between $v_i$ and $v_j$ is
\begin{eqnarray}
P_{ij}^{\mathrm{Dist}} = (\tilde{P}_{ij} + \tilde{P}_{ji}) / 2, \label{eq7}
\end{eqnarray}
where
\begin{eqnarray}
\tilde{P}_{ij} = \frac { k_ik_jf(d_{ij})} { \sum_{t=1}^n k_tf(d_{ti}) }. \label{eq8}
\end{eqnarray}
Here, $f: \mathbb{R}_{\geqslant0} \to [0,1]$ is a function which can be specified freely (we will explain the significance of $f$ later). $k_i$ denotes the degree of $v_i$ in the observed network. $d_{ij}$ denotes the distance between $v_i$ and $v_j$ in terms of $\rho$. Note that $d_{ij}$ is computed by a distance function between the attribute variables on $v_i$ and $v_j$, denoted by $\rho_i$ and $\rho_j$ \cite{LevandowskyDistanceBetweenSets}. Generally, $d_{ij}$ should satisfy the following constraints
\begin{eqnarray}
d_{ij} \geq 0 \mathrm{\ with\ equality\ IFF\ } \rho_i = \rho_j, \label{eq8-1}
\end{eqnarray}
\begin{eqnarray}
d_{ij} = d_{ji}. \label{eq8-2}
\end{eqnarray}

Our null model has the following properties. First, from Eq.~\eqref{eq7} we can find that
\begin{eqnarray}
P_{ij}^{\mathrm{Dist}} = P_{ji}^{\mathrm{Dist}}. \label{eq9}
\end{eqnarray}
This tells that links in our null model are undirected. Second, from Eq.~\eqref{eq8} we can derive that
\begin{eqnarray}
\sum_{i,j=1}^n \tilde{P}_{ij} = \sum_{i,j=1}^n \tilde{P}_{ji} = \sum_{i=1}^n k_i = 2m, \label{eq10}
\end{eqnarray}
and hence $\sum_{i,j=1}^n P_{ij}^{\mathrm{Dist}} = 2m$. This indicates that our null model preserves the number of links of the observed network. Third, from Eq.~\eqref{eq7} and \eqref{eq8} we can derive that
\begin{eqnarray}
k_{i}^{\mathrm{Dist}} = \sum_{j=1}^n P_{ij}^{\mathrm{Dist}} = \frac{k_i}{2} \Big[ 1+\sum_{j=1}^n \frac {k_jf(d_{ij})}{\sum_{t=1}^n k_tf(d_{tj})} \Big]. \label{eq11}
\end{eqnarray}
This implies that a node $v_i$ which has high degree in the observed network tends to have high degree in our null model. In addition, note that $k_{i}^{\mathrm{Dist}}$ is not necessarily equal to $k_i$. Thus, this may induce a slight shift in the distribution of $k_{i}^{\mathrm{Dist}}$, as compared to the distribution of $k_i$. Fourth, our null model can simulate the effect of $\rho$ on network topology by specifying appropriate function $f$. For example,
\begin{itemize}
\item If we specify $f(d)=e^{(-d)}$, two nodes which are similar have a higher chance of getting connected.
\item If we specify $f(d) =
\begin{cases}
1   & \text{if}\ d\leq\sigma \\
0   & \text{otherwise}
\end{cases}$, a node can only connect to those at a distance of no more than $\sigma$.
\item If we specify $f(d)=1$, $P_{ij}^{\mathrm{Dist}}$ is not related to $d_{ij}$, and our null model recovers the configuration model used in NG-Modularity.
\item If we specify $f(d)=e^{(-1/d)}$, two nodes which are dissimilar have a higher chance of getting connected --- this is actually a disassortativity effect \cite{JohnsonDisassortativity}.
\end{itemize}

%对于bb= 0 15 400 320
%第1个数字控制与左边的距离，越大与左边的距离越小
%第2个数字控制与下面的距离，越大与下边的距离越小
%第3个数字控制与右边的距离，越大与右边的距离越大，图片也越小
%第4个数字控制与上面的距离，越大与上面的距离越大
%\begin{comment}
\begin{figure}[!t]
\centering
\includegraphics[width=0.45\textwidth]{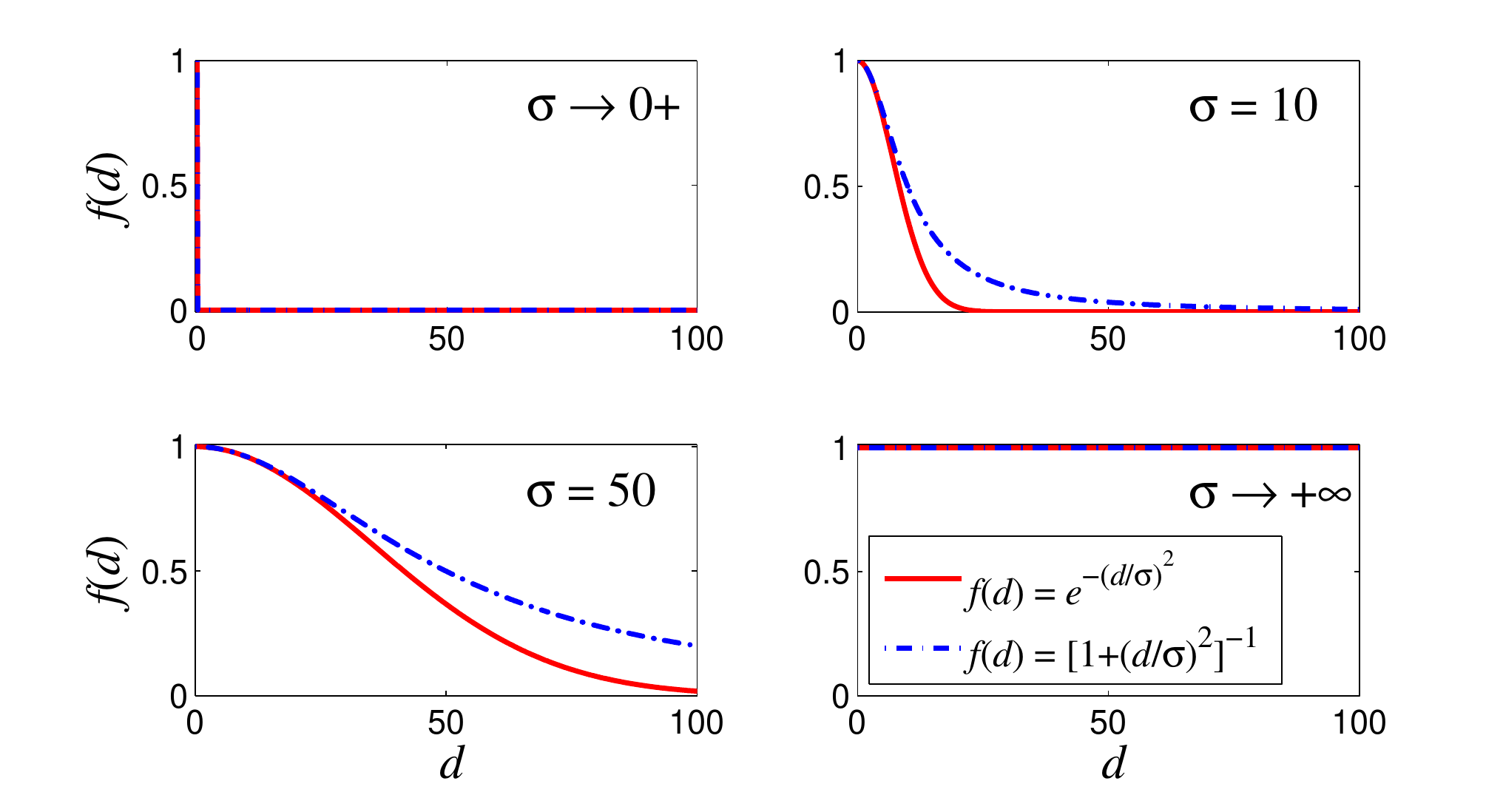}
\caption{\label{fig1} (Color online) The plot of function $f(d)=e^{-(d/\sigma)^2}$ and $f(d)=(1+(d/\sigma)^2)^{-1}$ at different values of $\sigma$.}
\end{figure}
%\end{comment}

\subsection{Dist-Modularity}\label{sec3.2}
Based on our null model, we define Dist-Modularity as
\begin{eqnarray}
\mathrm{Q}^{\mathrm{Dist}}(\mathcal{L})=\frac{1}{2m} \sum_{i,j=1}^n (A_{ij}-P_{ij}^{\mathrm{Dist}})\,\delta(l_{i},l_{j}). \label{eq12}
\end{eqnarray}
Note that we need to specify the function $f$ before using Dist-Modularity.

%$P_{il}^{\mathrm{Dist}}=\sum_{j=1}^{n}P_{ij}^{\mathrm{Dist}}\,\delta(l_{j},l)$
%compute the gains in $\mathrm{Q}^{\mathrm{Dist}}$ that would result from placing a node $v_i$ to its neighbor communities and place $v_i$ to the community which produces the highest gain.
%where $\bar{A}_{il}=\sum_{j=1}^{n}A_{ij}\,\delta(l_{j},l)$.
Like NG-Modularity, optimizing Dist-Modularity is NP-hard \cite{BrandesModularityNPCompleteness}. We can use heuristics such as Louvain algorithm \cite{BlondelFastUnfolding} and the advanced modularity-specialized label propagation algorithm (LPM+) \cite{LiuLPAmplus}, which were originally developed for optimizing NG-Modularity. The time complexity of Louvain and LPM+ algorithms can be analyzed as follows. First, we need to compute the denominator part of Eq.~\eqref{eq8} and keep the results for $i=1,\cdots,n$ in memory. This operation requires a complexity of $O(n^2)$. Second, suppose $c=|\{l_i | i=1,\cdots,n \}|$ is the number of communities. Also suppose we have enough memory to keep the $n \times c$ matrix $\bar{P}^{\mathrm{Dist}}$, whose elements are defined as $\bar{P}_{il}^{\mathrm{Dist}}=\sum_{j=1 \land j \ne i}^{n}P_{ij}^{\mathrm{Dist}}\,\delta(l_{j},l)$. At the initial stage where each node form a unique community, to save $\bar{P}^{\mathrm{Dist}}$ into memory requires a complexity of $O(n^2)$. Third, another computationally intensive step of these two algorithms is to move a node $v_i$ to a new community that would result in the highest gain in $\mathrm{Q}^{\mathrm{Dist}}$. We can derive that $v_i$'s new community membership can be computed as \cite{LiuLPAmplus}
\begin{eqnarray}
l_{i}^{\mathrm{new}} = \argmax_{l \in \{l_i\} \cup \{l_j|A_{ij}\ne0\}} (\sum_{j=1 \atop { j \ne i }}^{n}A_{ij}\,\delta(l_{j},l)-\bar{P}_{il}^{\mathrm{Dist}}). \label{eq12_5}
\end{eqnarray}
With $\bar{P}_{il}^{\mathrm{Dist}}$ kept in real time, this computation requires a complexity of $O(k_i)$. In addition, adjusting related elements of $\bar{P}^{\mathrm{Dist}}$ (at most $2k_i$ elements) due to the movement of $v_i$ requires another complexity of $O(k_i)$. Note that the node movement step is repeated sequentially for each node and iteratively until no gain in $\mathrm{Q}^{\mathrm{Dist}}$ can be attained. Suppose the number of iterations is $r$, which is a small number in practice. Then, the total time complexity of the two algorithms is near $O(n^2+rm)$.

%The computation of the gain has to be computed $k_i$ times for $v_i$, since there is at most $k_i$ communities in its neighborhood.
%$O(n^2)+O(pm) \approx O(n^2)$.

In the following, we discuss about how to detect communities beyond $\rho$ by Dist-Modularity. A key point to this problem is to find an appropriate function $f$ to simulate the assortativity effect of $\rho$ in the null model, so that such effect can be taken out when compared to the observed network. However, a challenge is that we do not know to what extent the network topology is affected by $\rho$. Thus, it is difficult to find a function $f$ directly.

%\begin{comment}
\begin{table*}[!t]
\begin{center}
\begin{tabular}{|l|l|l|}
\hline
               & \qquad\qquad\qquad\qquad\qquad\quad\ \,\;$\beta=0.3 $           & \qquad\qquad\qquad\qquad\qquad\quad\ \,\;$\beta =1.0$ \\
\hline
\multirow{3}{*}{$\epsilon=0.1$}    & \tabitem Space has the leading effect  & \tabitem Community membership has the leading effect \\
                                   & \tabitem Space and community membership are highly correlated & \tabitem Space and community membership are highly correlated \\
                                   & \tabitem Please see Fig.~\ref{fig2a} for visualization  & \tabitem Please see Fig.~\ref{fig2b} for visualization \\
\hline
\multirow{3}{*}{$\epsilon=0.5$}    & \tabitem Space has the leading effect  & \tabitem Community membership has the leading effect \\
                                   & \tabitem Space and community membership are uncorrelated & \tabitem Space and community membership are uncorrelated \\
                                   & \tabitem Please see Fig.~\ref{fig2c} for visualization  & \tabitem Please see Fig.~\ref{fig2d} for visualization \\
\hline
\end{tabular}
\end{center}
\caption{The four extreme cases in Cerina's synthetic spatial networks}
\label{table1}
\end{table*}
%\end{comment}

%对于bb= 0 15 400 320
%第1个数字控制与左边的距离，越大与左边的距离越小
%第2个数字控制与下面的距离，越大与下边的距离越小
%第3个数字控制与右边的距离，越大与右边的距离越大，图片也越小
%第4个数字控制与上面的距离，越大与上面的距离越大
%\begin{comment}
\begin{figure*}[!t]
\centering
\subfigure[][]{\label{fig2a}
\includegraphics[width=0.18\textwidth]{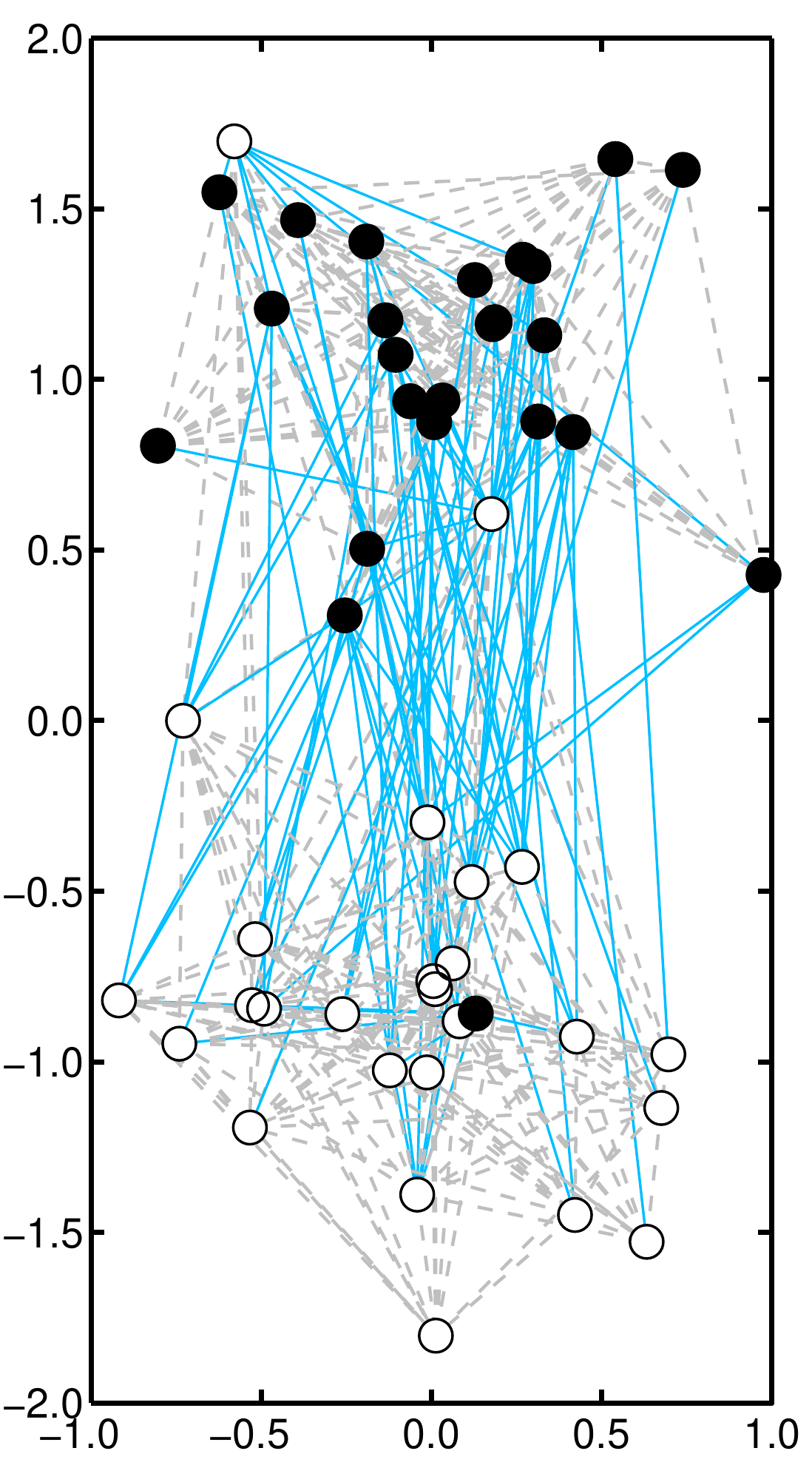}
}\,\,\,
\subfigure[][]{\label{fig2b}
\includegraphics[width=0.18\textwidth]{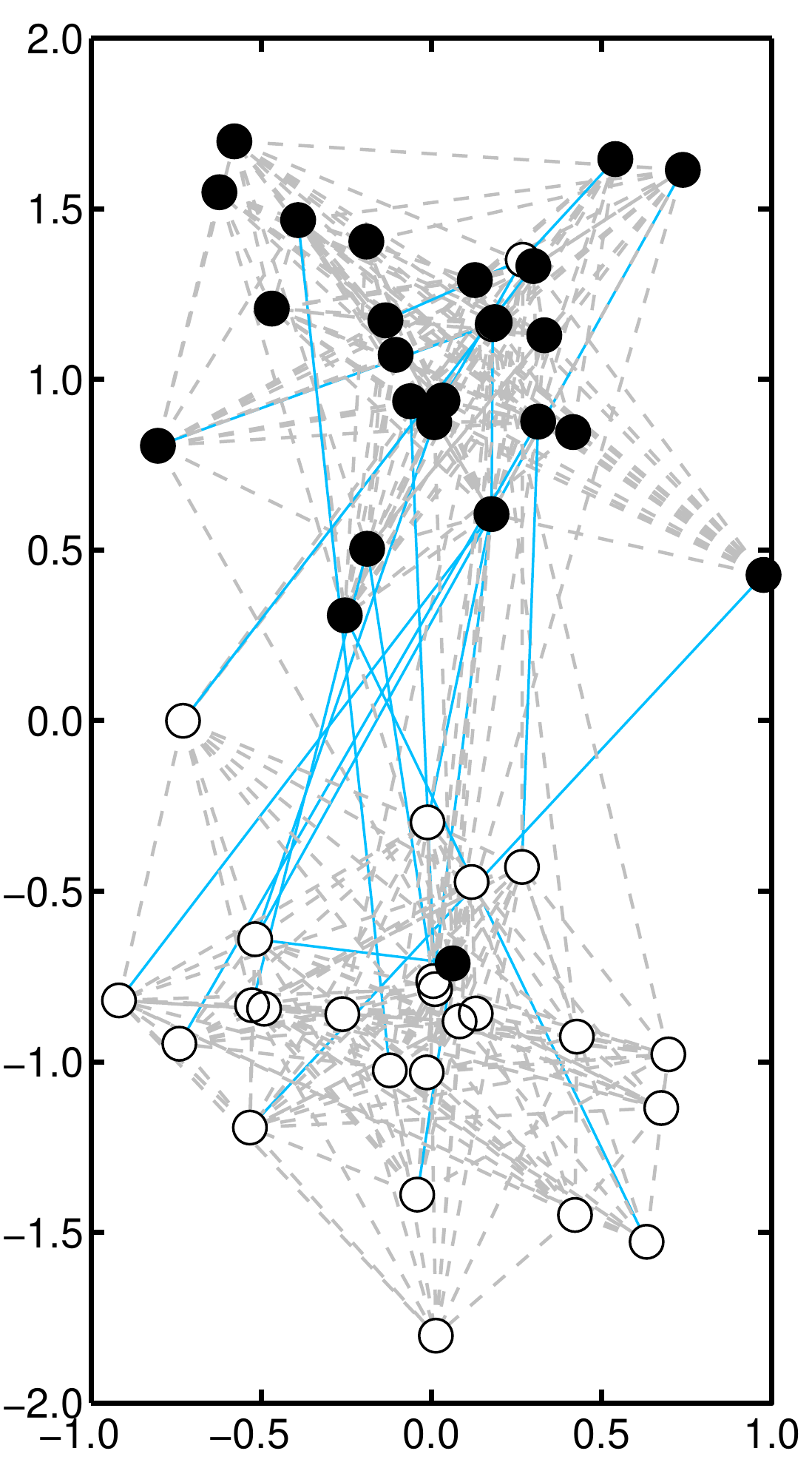}
}\,\,\,
\subfigure[][]{\label{fig2c}
\includegraphics[width=0.18\textwidth]{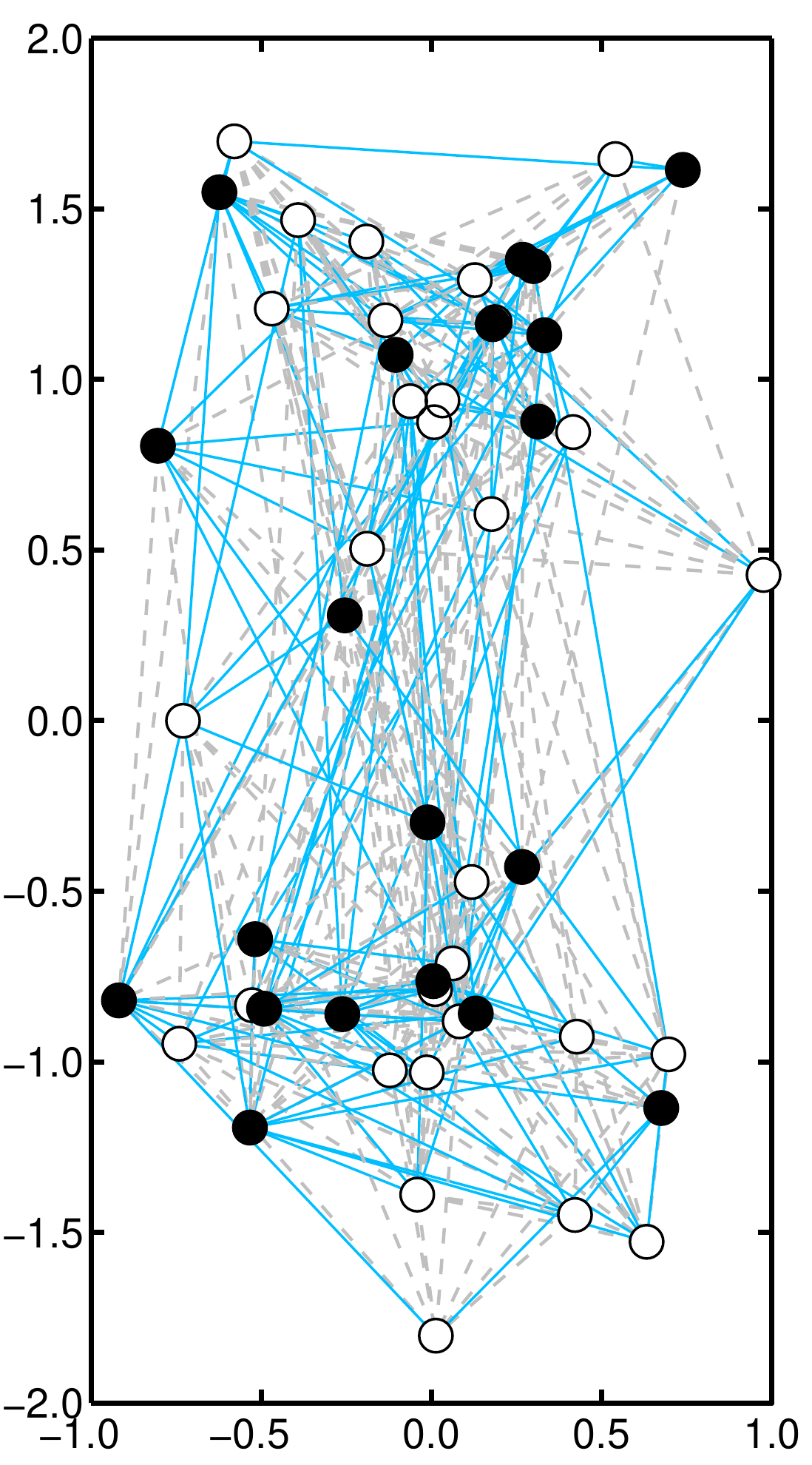}
}\,\,\,
\subfigure[][]{\label{fig2d}
\includegraphics[width=0.18\textwidth]{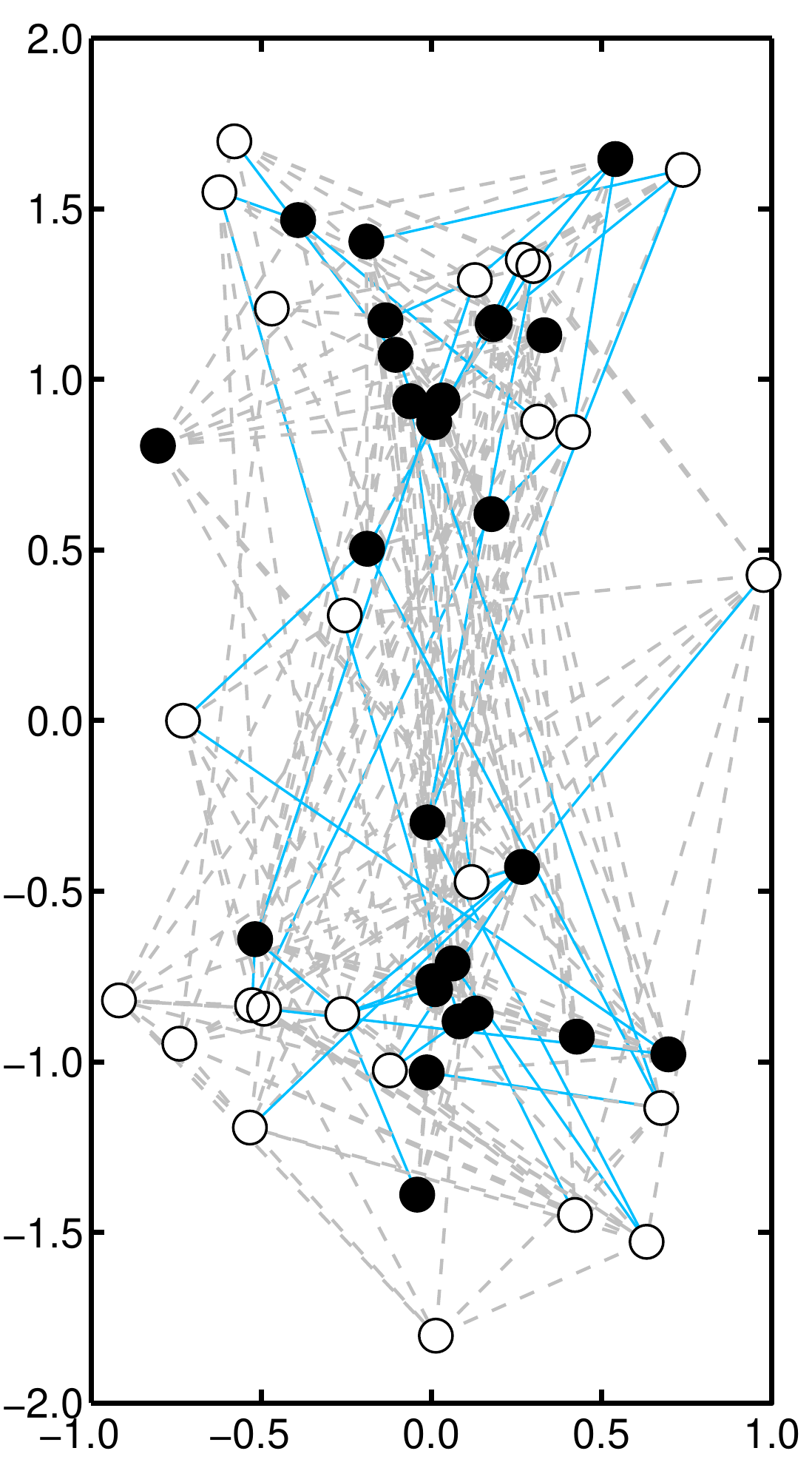}
}
\caption{\label{fig2} (Color online) Visualization of Cerina's synthetic spatial networks at the four extreme cases: (a) $\epsilon=0.1$, $\beta=0.3$, (b) $\epsilon=0.1$, $\beta=1.0$, (c) $\epsilon=0.5$, $\beta=0.3$, (d) $\epsilon=0.5$, $\beta=1.0$. For the sake of clarity, only 50 out of the 100 nodes are displayed here. The nodes in community $C_{+1}$ and $C_{-1}$ are painted in black and white, respectively. The between-community and within-community links are plotted by solid and dashed lines, respectively.}
\end{figure*}
%\end{comment}

In the framework of Dist-Modularity, our procedure contains three steps. First, we choose a distance function \cite{LevandowskyDistanceBetweenSets} to compute $d_{ij}$. Second, we probe $f$ using parameterized functions such as $f(d)=e^{-(d/\sigma)^2}$ and $f(d)=(1+(d/\sigma)^2)^{-1}$, where $\sigma\in(0,+\infty)$ is a parameter \footnote{Note that there are many candidate functions. Choosing the right function form should depend on the background knowledge of the problem we are dealing with. Please see Section~\ref{sec4} for more information.}. A benefit of these functions is that we can tune the parameter to simulate the assortativity effect at different degrees. Take the function $f(d)=e^{-(d/\sigma)^2}$ as an example. At the extreme of $\sigma \to 0+$, the null model has the strongest assortativity effect and a node $v_i$ can only connect to its most similar nodes $v_j$ satisfying $d_{ij}=0$ . At the extreme of $\sigma \to +\infty$, we can derive
\begin{eqnarray}
\lim\limits_{\sigma \to +\infty} \big( P_{ij}^{\mathrm{Dist}} | f(d)=e^{-(d/\sigma)^2} \big) = k_ik_j/2m. \label{eq14}
\end{eqnarray}
That is, the null model has no assortativity effect. As $\sigma$ increases from $0$ to $+\infty$ (see Fig.~\ref{fig1}), the assortativity effect in the null model gradually fades.

In the third step, we optimize Dist-Modularity at various estimated values of the parameter and select one that brings the ``best'' partition. A possible way is the alternative parameter selection method proposed by Expert et al. \cite{ExpertSpatialCommunity}, which seeks to find a consensus partition. More precisely, suppose $\mathcal{L}_{\sigma_i}$ is a partition obtained by optimizing Dist-Modularity at $\sigma=\sigma_i (i = 1,\cdots,s)$, which are possible values of the parameter. We compute the average normalized mutual information (NMI) \cite{FredNMI,DanonCompareCommunityAlgo} as
\begin{eqnarray}
I^{\mathrm{avg}}(\mathcal{L}_{\sigma_i}) = \sum_{j=1 \atop { j \ne i }}^{s}I(\mathcal{L}_{\sigma_i}, \mathcal{L}_{\sigma_j}), \label{eq15}
\end{eqnarray}
where $I$ represents the function of NMI. The partition with the highest $I^{\mathrm{avg}}$ score, which is the consensus partition and the closest to the others, is our final partition for communities beyond $\rho$.

The time complexity of the above procedure is as follows. First, optimizing Dist-Modularity at $\sigma=\sigma_1,\cdots,\sigma_s$ requires a complexity of $O(s(n^2+rm))$. Second, computing the $I^{avg}$ score for $s$ partitions requires a complexity of $O(s^{2}(n+c^2))$. Finally, the total complexity of detecting communities beyond $\rho$ by Dist-Modularity is $O(s(n^2+rm))+O(s^{2}(n+c^2))$. Suppose $s,c,r \ll n$, and the network is sparse such that $O(m)=O(n)$. This complexity can be simplified to $O(sn^2)$.

\section{Experiments}\label{sec4}
Modularity is widely used for detecting communities. In general, one takes Modularity as an objective function and finds the ``best'' community partition by an optimization algorithm \cite{NewmanGreedy,ClausetFastGreedy,NewmanModularityEigenvectors,DuchExtremalOptimization,MedusSA,WakitaCommunityMegaScaleNetwork,SchuetzMultistepGreedy2,BarberLPAm}. In this section, we use three examples to demonstrate that Dist-Modularity is practically useful in detecting communities beyond assortativity-related attributes. We also compare Dist-Modularity with NG-Modularity and Spa-Modularity. All results are obtained using LPM+ optimization algorithm, because it can find higher Modularity scores than Louvain algorithm, without much additional running time \cite{LiuLPAmplus}.

\subsection{Cerina's Synthetic Spatial Networks}\label{sec4.1}
The first example is the synthetic spatial networks proposed by Cerina et al. \cite{CerinaSpatialCorrelation} for testing whether a Modularity can detect communities beyond the space attribute. Our scheme is as follows. 1) We generate a set of synthetic spatial networks with known community structure (the true partition). The network topology is generated based on both space and community membership --- two nodes which are spatially closer have a higher chance of getting connected, and two nodes which have the same community membership also have a higher chance of getting connected. 2) We apply NG-Modularity, Spa-Modularity, and Dist-Modularity to these networks to detect the communities beyond space. 3) We compute NMI between the true partition and partitions by the three Modularities. The higher the NMI score, the better the corresponding Modularity.

The network generation procedure contains the following three steps.
\begin{enumerate}[(1)]
\item Generate 100 nodes in a $(x,y)$ 2D-space. The first 50 nodes are around the North center $(0,1)$ and fall in the North area $\{(x,y)|-1<x<1, 0<y<2\}$. The second 50 nodes are around the South center $(0,-1)$ and fall in the South area $\{(x,y)|-1<x<1, -2<y<0\}$. We generate the coordinates $(x_i,y_i)$ of a node $v_i$ according to probability $p_{\mathrm{coord}}(x_i,y_i) \propto e^{(-d_{ic})}$, where $d_{ic}$ is the Euclidean distance between $v_i$ and its corresponding center.
\item Arrange nodes into communities $C_{+1}$ and $C_{-1}$. We assign the community membership $l_i$ of node $v_i$ as
\begin{eqnarray}
l_i =
\begin{cases}
-\mathrm{sgn}(y_i)  &  \text{with probability}\ \epsilon\\
+\mathrm{sgn}(y_i)  &  \text{with probability}\ 1-\epsilon,
\end{cases}
\label{eq15.5}
\end{eqnarray}
where $\mathrm{sgn}$ denotes the sign function, and $\epsilon \in [0.1,0.5]$ is a parameter representing the correlation between space and community membership. In the case $\epsilon = 0.1$, space and community membership are highly correlated, such that $90\%$ of the North nodes are assigned to community $C_{+1}$ and $90\%$ of the South nodes to community $C_{-1}$. In the case $\epsilon = 0.5$, space and community membership are totally uncorrelated, and nodes are assigned to either communities with probability 0.5.
\item Generate links. We generate a link between $v_i$ and $v_j$ according to probability $p_{\mathrm{link}}(v_i,v_j) \propto e^{\beta l_il_j-d_{ij}}$, where $d_{ij}$ is the Euclidean distance between $v_i$ and $v_j$. We can see that $p_{\mathrm{link}}(v_i,v_j)$ is positively related to $l_il_j$ and $d_{ij}$. Thus, the existence of a link is affected by both space and community membership. Here $\beta \in [0.3,1.0]$ is a parameter determining space, or community membership, which has the leading effect in network topology. In the case $\beta = 0.3$, space has the leading effect and links are essentially between spatially close nodes. In the case $\beta = 1.0$, community membership has the leading effect and links are essentially between nodes of the same community.
\end{enumerate}

%对于bb= 0 15 400 320
%第1个数字控制与左边的距离，越大与左边的距离越小
%第2个数字控制与下面的距离，越大与下边的距离越小
%第3个数字控制与右边的距离，越大与右边的距离越大，图片也越小
%第4个数字控制与上面的距离，越大与上面的距离越大
%\begin{comment}
\begin{figure*}[!t]
\centering
\includegraphics[width=0.31\textwidth]{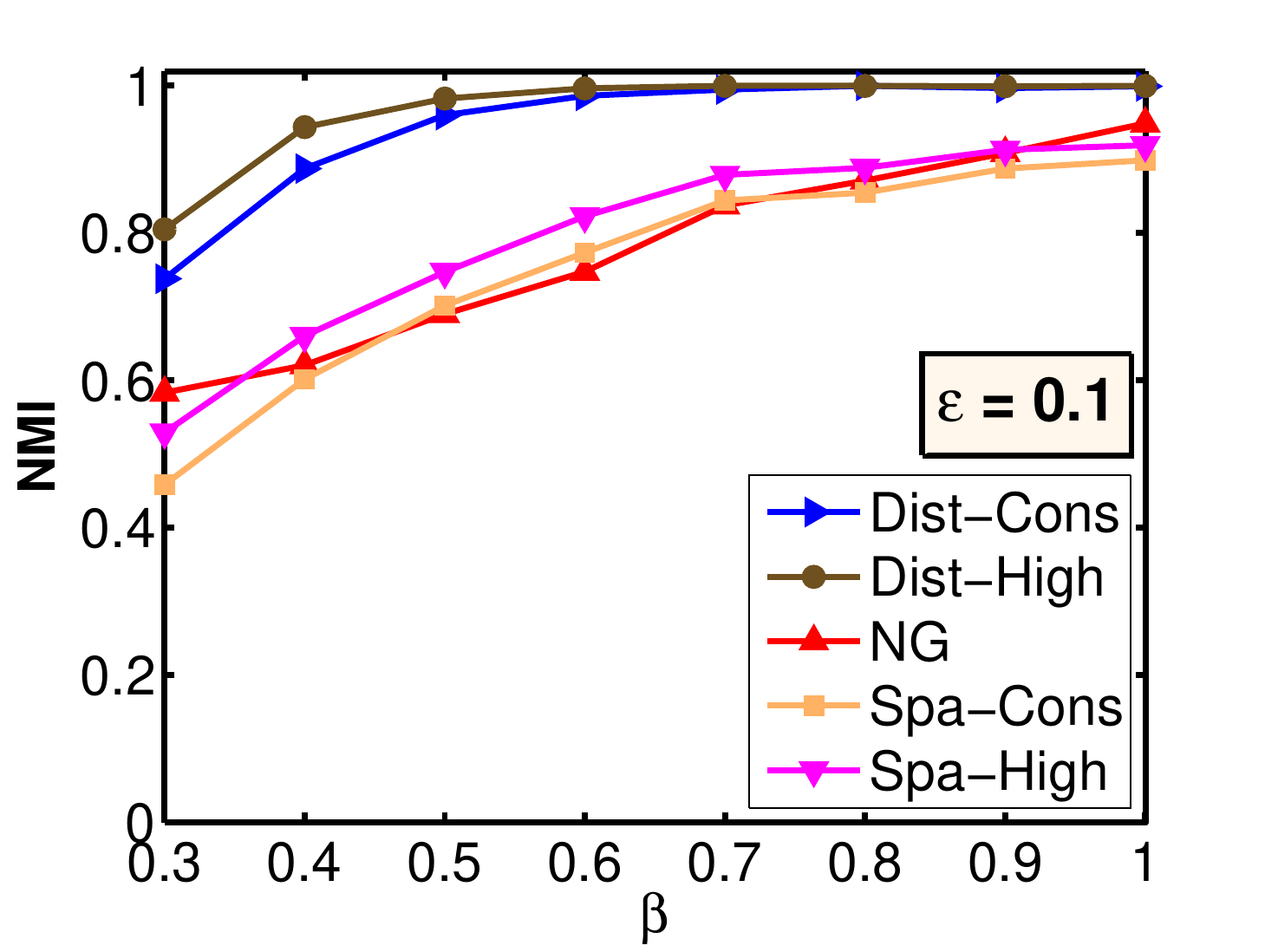}
\includegraphics[width=0.31\textwidth]{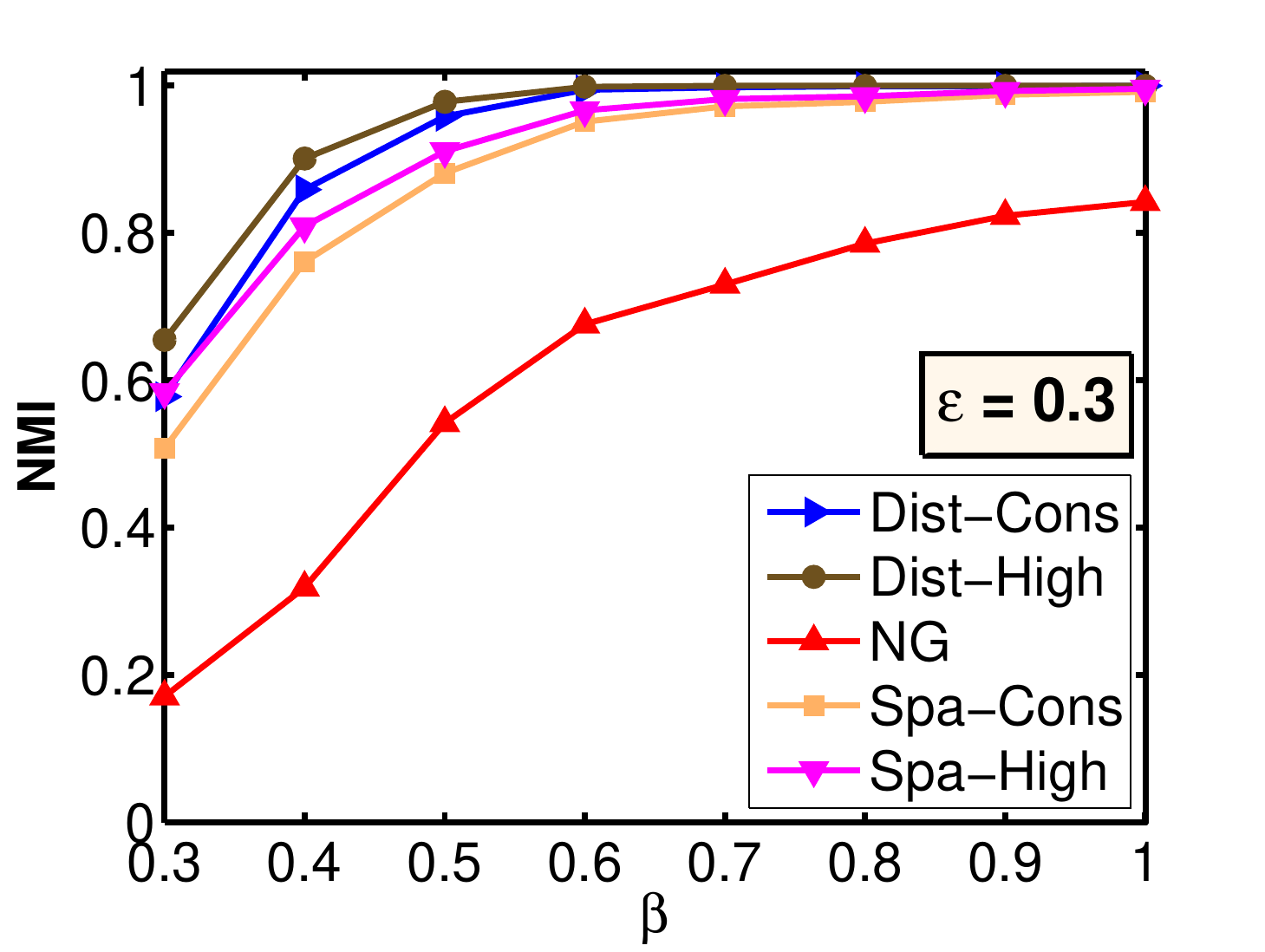}
\includegraphics[width=0.31\textwidth]{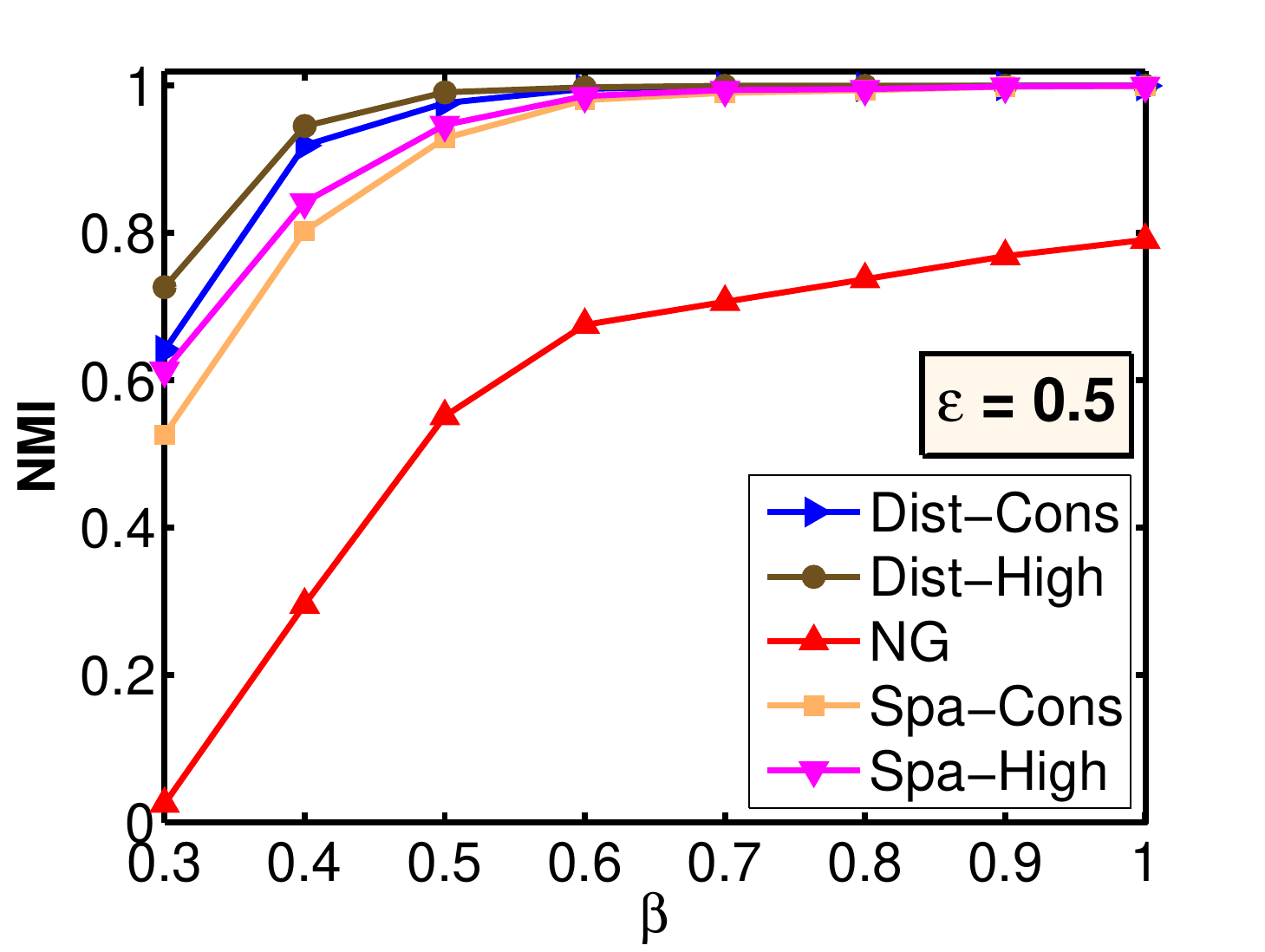}
\caption{\label{fig3} (Color online) The NMI scores by NG-Modularity, Spa-Modularity, and Dist-Modularity in Cerina's synthetic spatial networks, for $\beta \in [0.3,1.0]$, $\epsilon = $0.1, 0.3, and 0.5. Each score is based on an average of 100 implementations. NG stands for scores by NG-Modularity. Spa-High and Spa-Cons stand for the highest scores and the scores of the consensus partitions by Spa-Modularity, respectively. Dist-High and Dist-Cons stand for the highest scores and the scores of the consensus partitions by Dist-Modularity, respectively.}
\end{figure*}
%\end{comment}

By tuning parameters $\epsilon$ and $\beta$, we can create various cases reflecting the interplay between space and community membership. In particular, we illustrate four extreme cases of this series of networks in Table~\ref{table1} and Fig.~\ref{fig2}. These networks enable us to systematically study the performance of different Modularitites.

We adopted the following procedure to apply Dist-Modularity. First, we used the Euclidean distance between $(x_i,y_i)$ and $(x_j,y_j)$ to compute $d_{ij}$. Second, considering that we did not know to what extent the network topology is affected by space, we specified $f(d)=e^{-(d/\sigma)^2}$, so that we can tune $\sigma$ to simulate the space effect at different degrees. Third, suppose $\bar{d}=\sum_{i,j=1}^{n}d_{ij}/n^2$ is the average distance of all node pairs. We conducted the alternative parameter selection for $\sigma \in [0.1\bar{d},2.0\bar{d}]$, with a step length of $0.1\bar{d}$. Then, we reported the NMI score of the consensus partition. In addition, we reported the highest NMI score obtained in this $\sigma$ interval.

%\footnote{The reason we chose this interval is that it covers the most possible space effect.}

As for Spa-Modularity, Expert et al. suggested a binning strategy to smoothen the probability function $p(d)$ which is expressed in Eq.~\eqref{eq5} \cite{ExpertSpatialCommunity}. More precisely, they introduced a bin size parameter $\tau$, which can influence the form of $p(d)$. For various possible values of $\tau$, they conducted the alternative parameter selection. We followed their suggestions in applying Spa-Modularity, and reported the NMI score of the consensus partition. We also reported the highest NMI score obtained by various values of $\tau$.

Fig.~\ref{fig3} shows the NMI scores by NG-Modularity, Spa-Modularity, and Dist-Modularity, for $\beta \in [0.3,1.0]$, $\epsilon = $0.1, 0.3, and 0.5. Overall, the NMI scores follow an upward trend as $\beta$ increases from 0.3 to 1.0. This is because, as $\beta$ increases, the community membership has greater effect in network topology, and there are more links between nodes of the same community. As a result, it becomes easier to detect the communities.

%对于bb= 0 15 400 320
%第1个数字控制与左边的距离，越大与左边的距离越小
%第2个数字控制与下面的距离，越大与下边的距离越小
%第3个数字控制与右边的距离，越大与右边的距离越大，图片也越小
%第4个数字控制与上面的距离，越大与上面的距离越大
%\begin{comment}
\begin{figure*}[!t]
\centering
\subfigure[][]{\label{fig4a}
\includegraphics[width=0.27\textwidth]{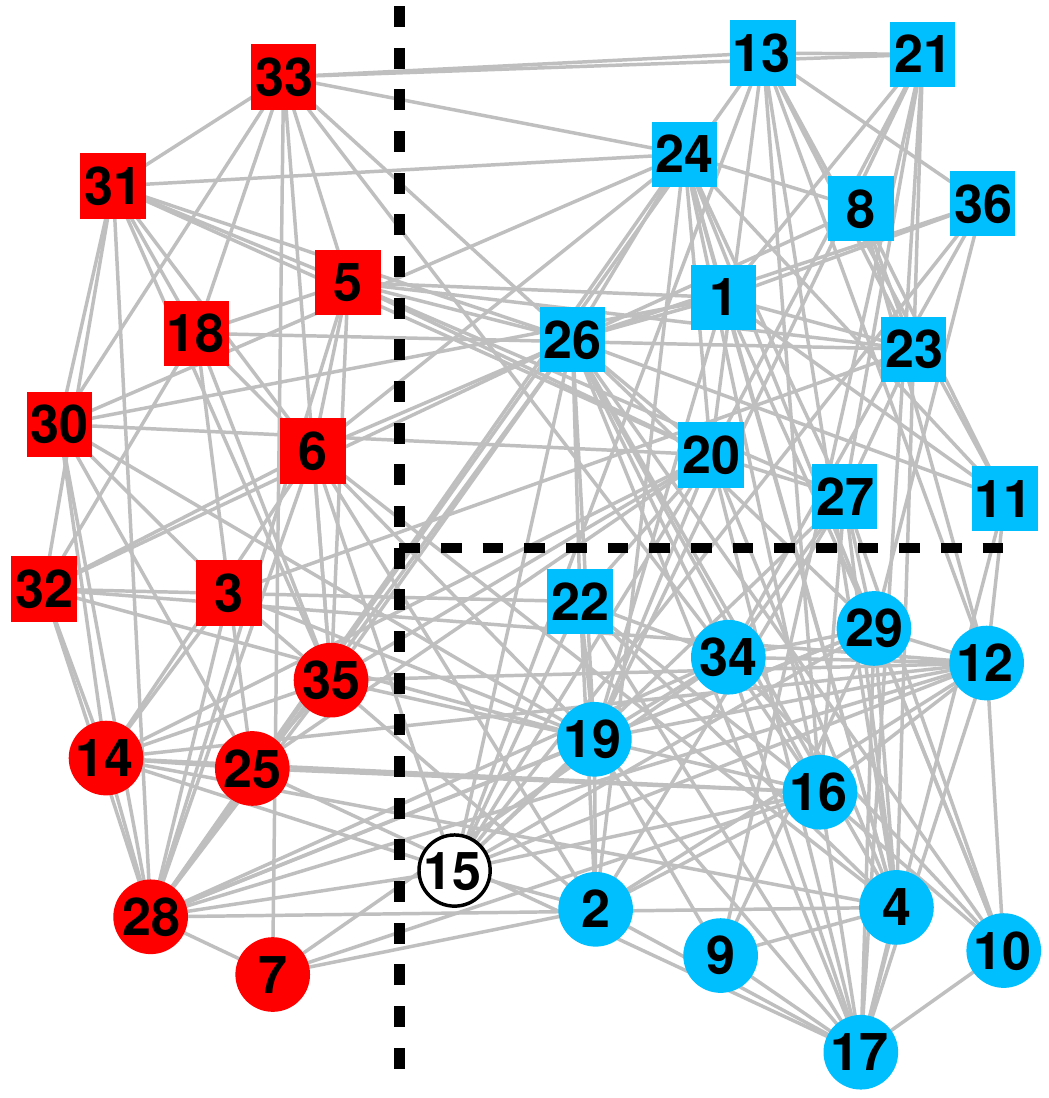}
}\qquad
\subfigure[][]{\label{fig4b}
\includegraphics[width=0.27\textwidth]{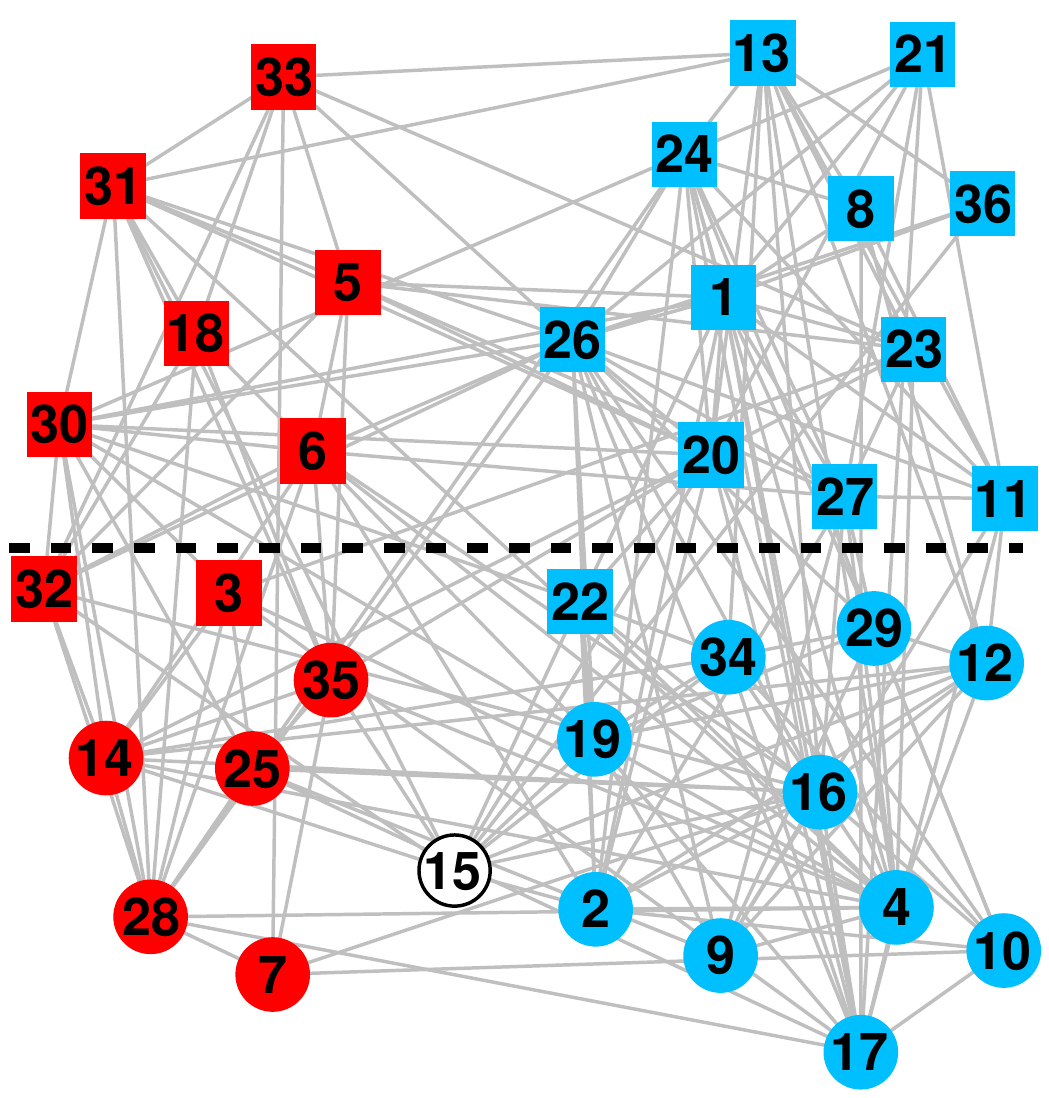}
}\qquad
\subfigure[][]{\label{fig4c}
\includegraphics[width=0.27\textwidth]{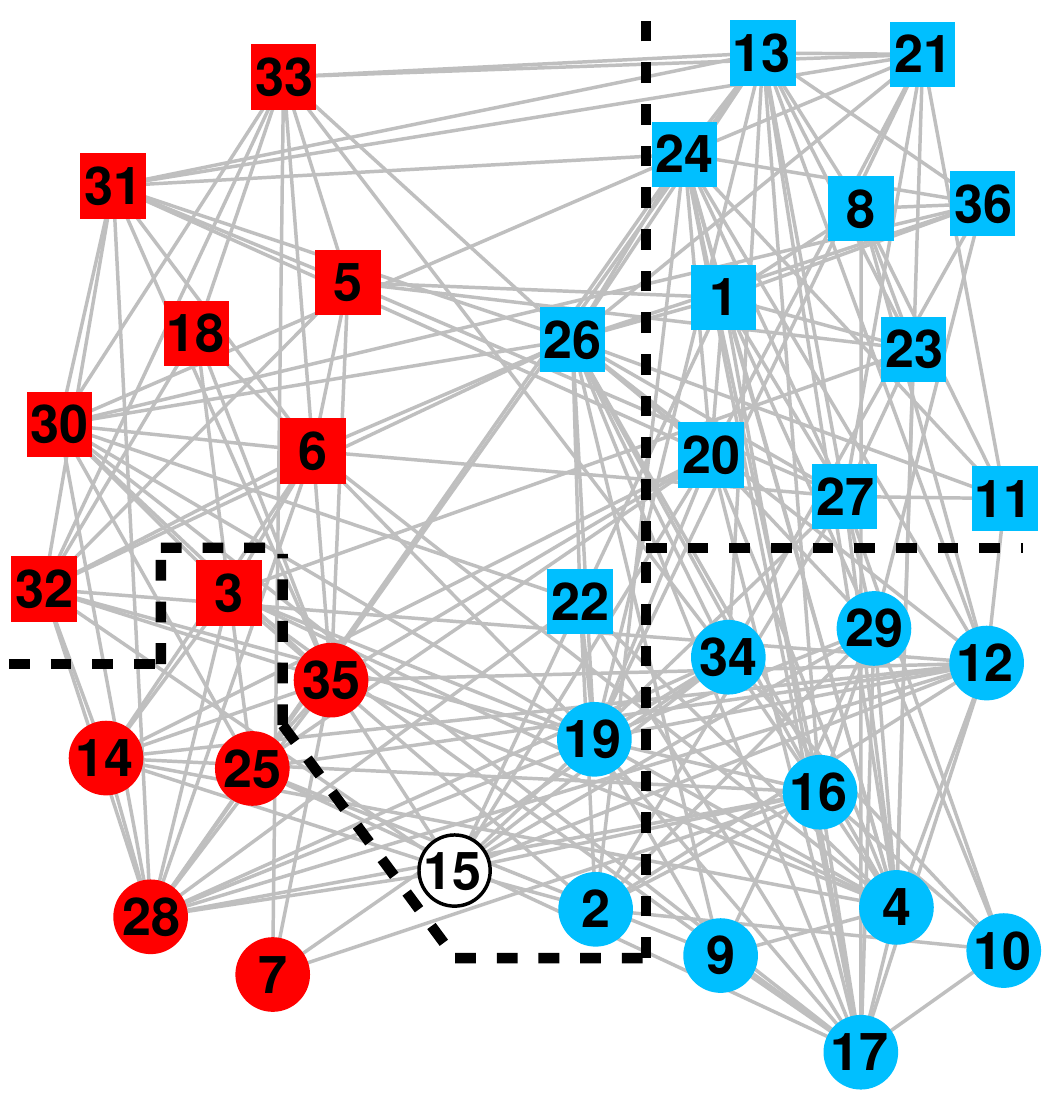}
}
\caption{\label{fig4} (Color online) Visualization of Lazega's partner advice network. Partners whose practice is litigation and corporate law are symbolized as square and round nodes, respectively. Partners whose office location is Hartford, Providence, and Boston are painted in red (dark gray), white, and blue (light gray), respectively. (a) The Partition by NG-Modularity. (b) The Partition by Dist-Modularity. (c) The Partition by Spa-Modularity.}
\end{figure*}
%\end{comment}

By comparison, Dist-Modularity performs the best, followed by Spa-Modularity. The reason for Spa-Modularity's inferiority is because that it failed to accurately simulate the space effect in its null model. To see it, let us look back at Eq.~\eqref{eq6}, the foundation of Spa-Modularity null model. This null model tries to simulate the space effect, so that the number of links between nodes at distance $d$ is the same as that number in the observed network. However, since network topology is affected by both space and community membership, the pure effect caused by space itself is not what we observed in the network topology \footnote{Actually, under the space effect itself, the number of links between nodes at distance $d$ can be larger or smaller than that number in the observed network. This depends on two factors: 1) the proportion of effects by space and community membership. 2) the correlation between space and community membership.}. In particular, when there is a strong correlation between space and community membership, the effect by space is quite different from what we observed in the network topology. For this reason, Spa-Modularity does not perform well when $\epsilon = 0.1$.

On the other hand, NG-Modularity cannot detect the true communities with $100\%$ accuracy, even when $\beta$ is large. This is because that NG-Modularity does not take attributes into account, and thus cannot take out the space effect.

Furthermore, we can find that the gap between the highest NMI scores and scores of the consensus partitions by Dist-Modularity is not large. This indicates that the alternative parameter selection can help us find a good partition which is close to the best possible one. In real-world applications, we do not know the true partition. Thus, this gap also implies that there is still a potential room for improvement --- If we can reduce the search interval of $\sigma$ based on some background knowledge of the network, our results can be even better.

\subsection{Lazega's Partner Advice Network}\label{sec4.2}
The second example is based on a dataset collected by Lazega on relations between partners in a New England law firm \cite{LazegaLawyers,SnijdersExponentialRandomGraphModel}. From the dataset, we constructed a symmetrized network, where nodes represent 36 partners in the firm, and links represent 395 advisee-adviser relations (we ignore the direction of links). We weight a link by 1 if one partner has ever sought professional advices from the other, and weight a link by 2 if both partners have sought advices from each other.

Moreover, various partners' attributes are also part of the dataset. For example, we have information about age, gender, office location (Hartford, Providence, and Boston), and practice (litigation or corporate law) of each partner. In Fig.~\ref{fig4}, we use red (dark gray), white, and blue (light gray) colors to differentiate partners in terms of office location, and use square and round symbols to differentiate partners in terms of practice.

%对于bb= 0 15 400 320
%第1个数字控制与左边的距离，越大与左边的距离越小
%第2个数字控制与下面的距离，越大与下边的距离越小
%第3个数字控制与右边的距离，越大与右边的距离越大，图片也越小
%第4个数字控制与上面的距离，越大与上面的距离越大
%\begin{comment}
\begin{figure}[!t]
\centering
\subfigure[][]{\label{fig5a}
\includegraphics[width=0.21\textwidth]{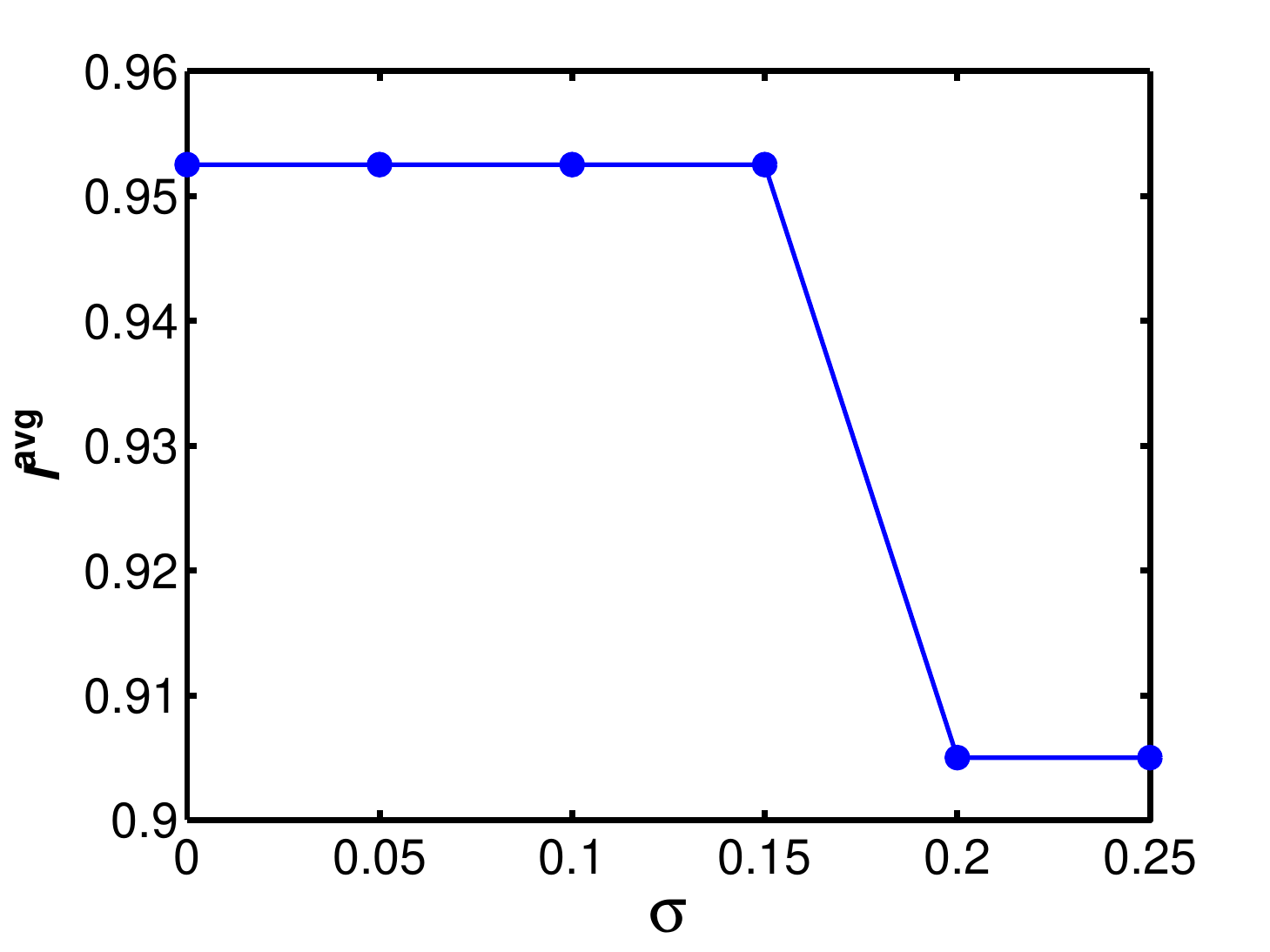}
}
\subfigure[][]{\label{fig5b}
\includegraphics[width=0.21\textwidth]{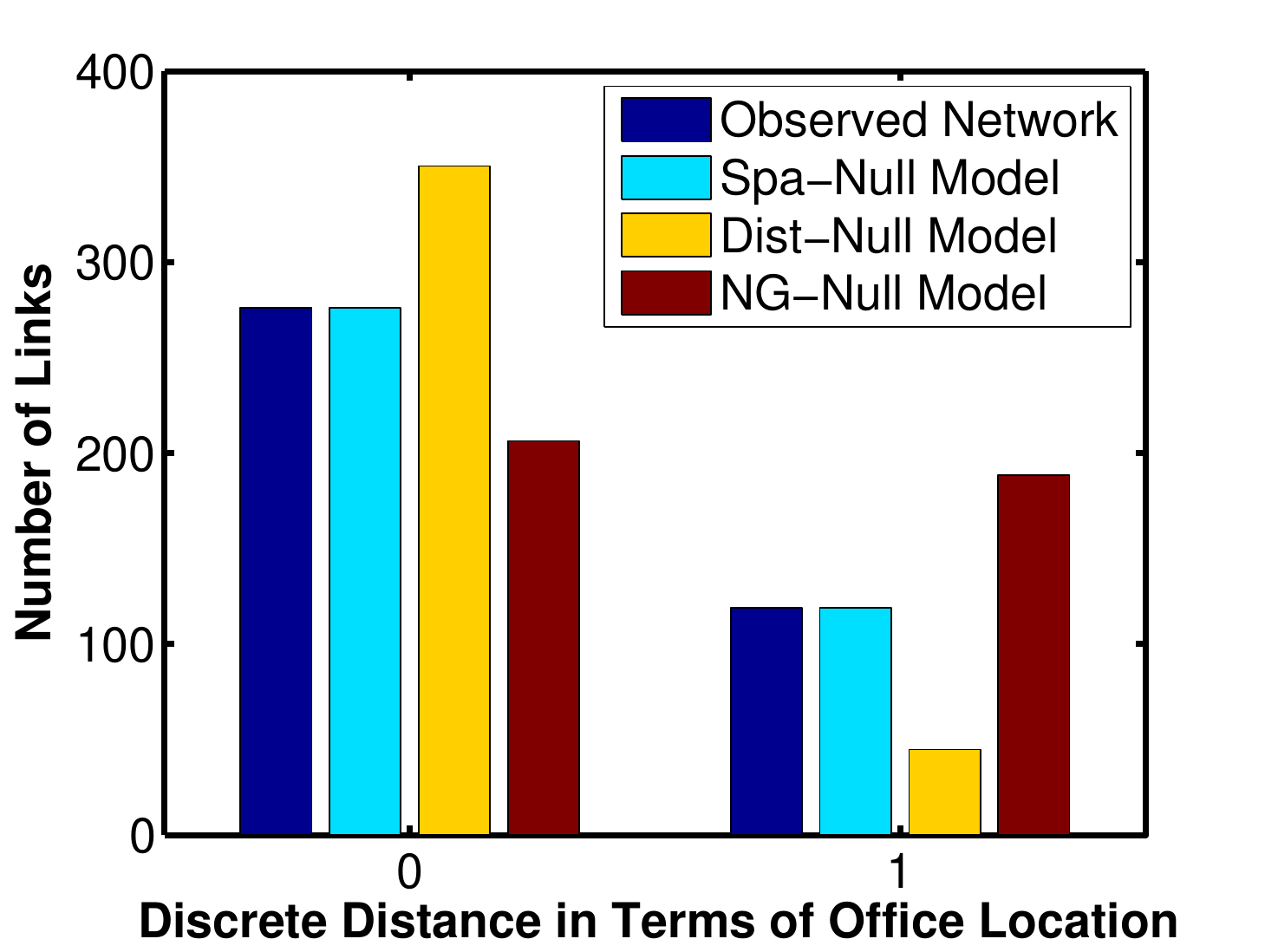}
}
\caption{\label{fig5} (Color online) Lazega's partner advice network. (a) The averaged normalized mutual information as a function of $\sigma$. (b) The effects of office location in the observed network, the Spa-Modularity null model, the Dist-Modularity null model ($\sigma=0.1)$, and the NG-Modularity null model.
}
\end{figure}
%\end{comment}

We can expect that the attributes of office location and practice have significant assortativity effects on network topology, because the advisee-adviser relationship are more likely between partners working in the same office and those with the same practice. Fig.~\ref{fig4a} shows the three-community partition by NG-Modularity. Note that this partition successfully separates red nodes (office = Hartford) from white and blue nodes (office = Providence and office = Boston), and it also separates square nodes (practice = litigation) from round nodes (practice = corporate law) to some degree. Thus, without taking attributes into account, NG-Modularity brings a compromise between the partition based on office location and the partition based on practice. An interesting problem is can we find the communities based on practice beyond the attribute of office location.

To solve this problem by Dist-Modularity, we used the discrete distance between office locations of $v_i$ and $v_j$ to compute $d_{ij}$ . That is
\begin{eqnarray}
d_{ij} =
\begin{cases}
0   & \text{if}\ v_i \text{ and } v_j \text{ has the same office location};\text{\ \ \ \ }\\
1   & \text{otherwise}.
\end{cases}
\label{eq16}
\end{eqnarray}
Considering that the link probability is much different for partners working in the same and different offices, we specified function $f$ as
\begin{eqnarray}
f(d) =
\begin{cases}
1   & \text{if}\ d=0;\\
\sigma   & \text{otherwise},
\end{cases}
\label{eq17}
\end{eqnarray}
where $\sigma$ is a parameter representing the probability of a link that exist between two nodes with different office locations in the Dist-Modularity null model. Then, we conducted the alternative parameter selection for $\sigma \in [0,0.25]$, with a step length of 0.05. The consensus partition was obtained at $\sigma \in [0,0.15]$, as shown in Fig.~\ref{fig5a}. Fig.~\ref{fig4b} illustrates this two-community partition. We can find that it is almost the same as the partition based on practice, with only three nodes ($\#3$, $\#22$, and $\#32$) classified differently. Compared to the three-community partition by Spa-Modularity, as shown in Fig.~\ref{fig4c}, our partition is much closer to the partition based on practice.

%对于bb= 0 15 400 320
%第1个数字控制与左边的距离，越大与左边的距离越小
%第2个数字控制与下面的距离，越大与下边的距离越小
%第3个数字控制与右边的距离，越大与右边的距离越大，图片也越小
%第4个数字控制与上面的距离，越大与上面的距离越大
%\begin{comment}
\begin{figure}[!t]
\centering
\subfigure[][]{\label{fig6a}
\includegraphics[width=0.21\textwidth]{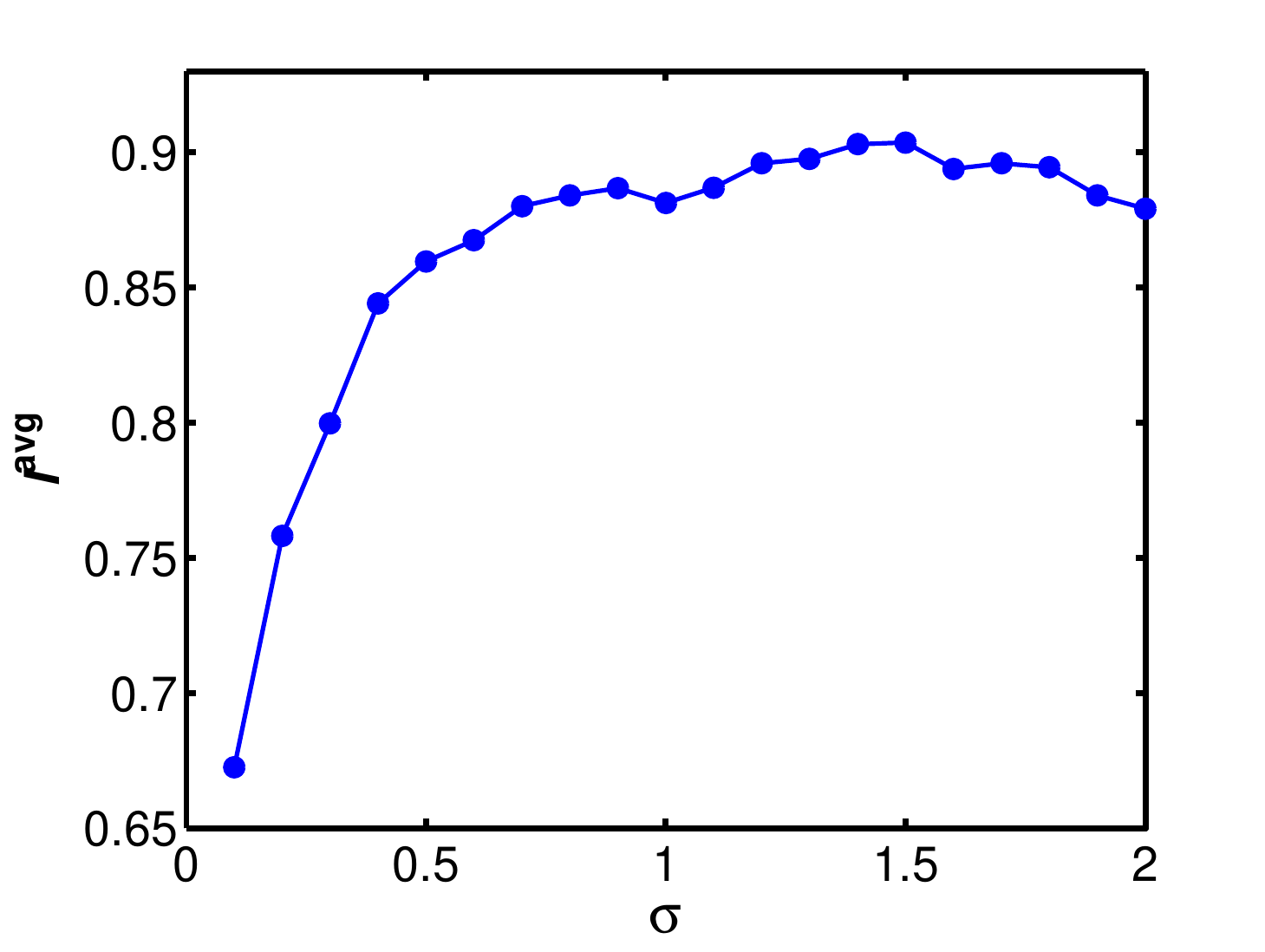}
}
\subfigure[][]{\label{fig6b}
\includegraphics[width=0.21\textwidth]{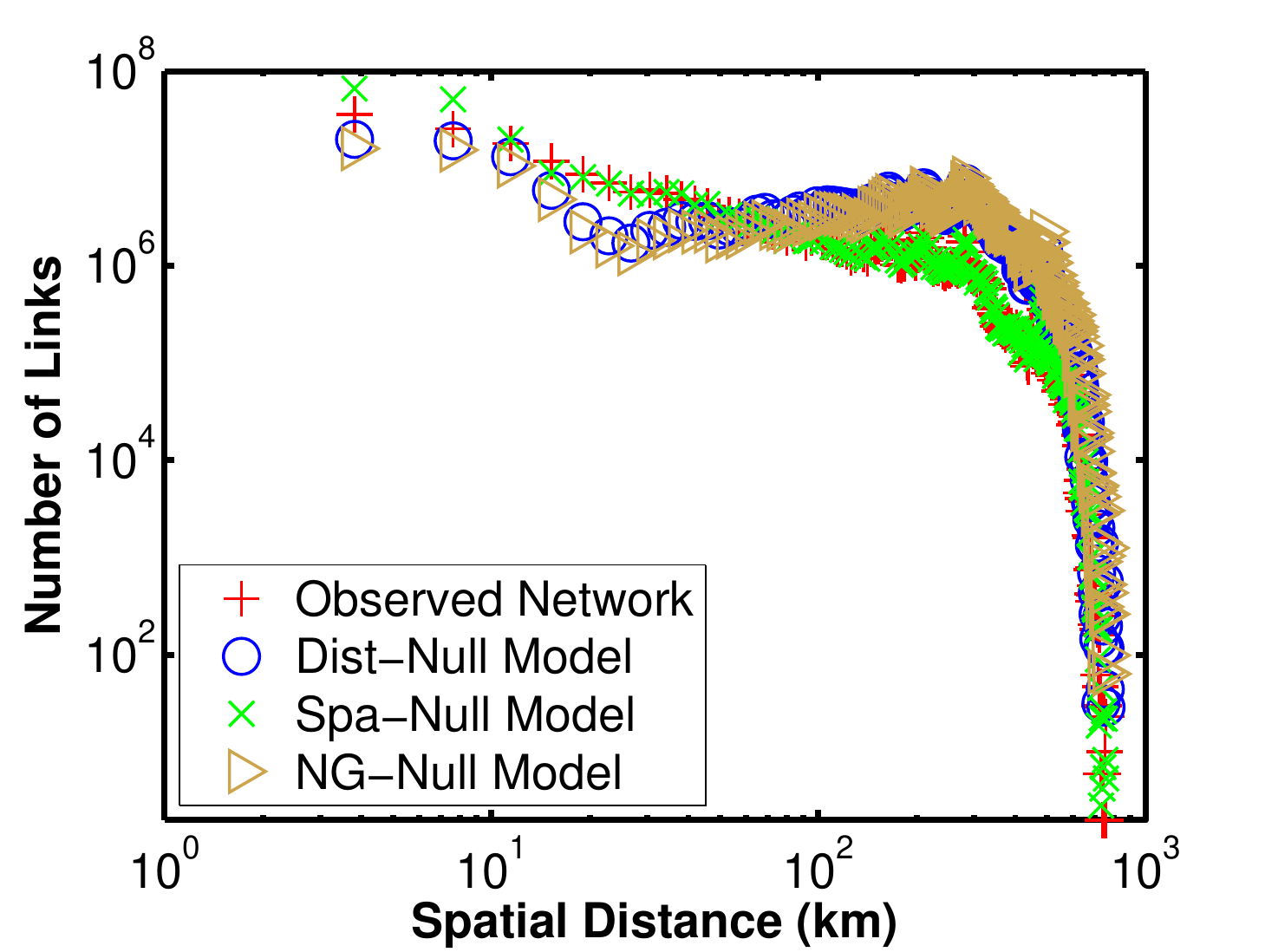}
}
\caption{\label{fig6} (Color online) D4D Antenna Network. (a) The averaged normalized mutual information as a function of $\sigma$. (b) The space effect in the observed network, the Spa-Modularity null model, the Dist-Modularity null model, and the NG-Modularity null model.
}
\end{figure}
%\end{comment}

%\begin{comment}
\begin{figure*}[!t]
\centering
\subfigure[][]{\label{fig7a}
\setlength{\abovecaptionskip}{0pt}%设置图形与标题之间的距离为0
\includegraphics[width=0.39\textwidth]{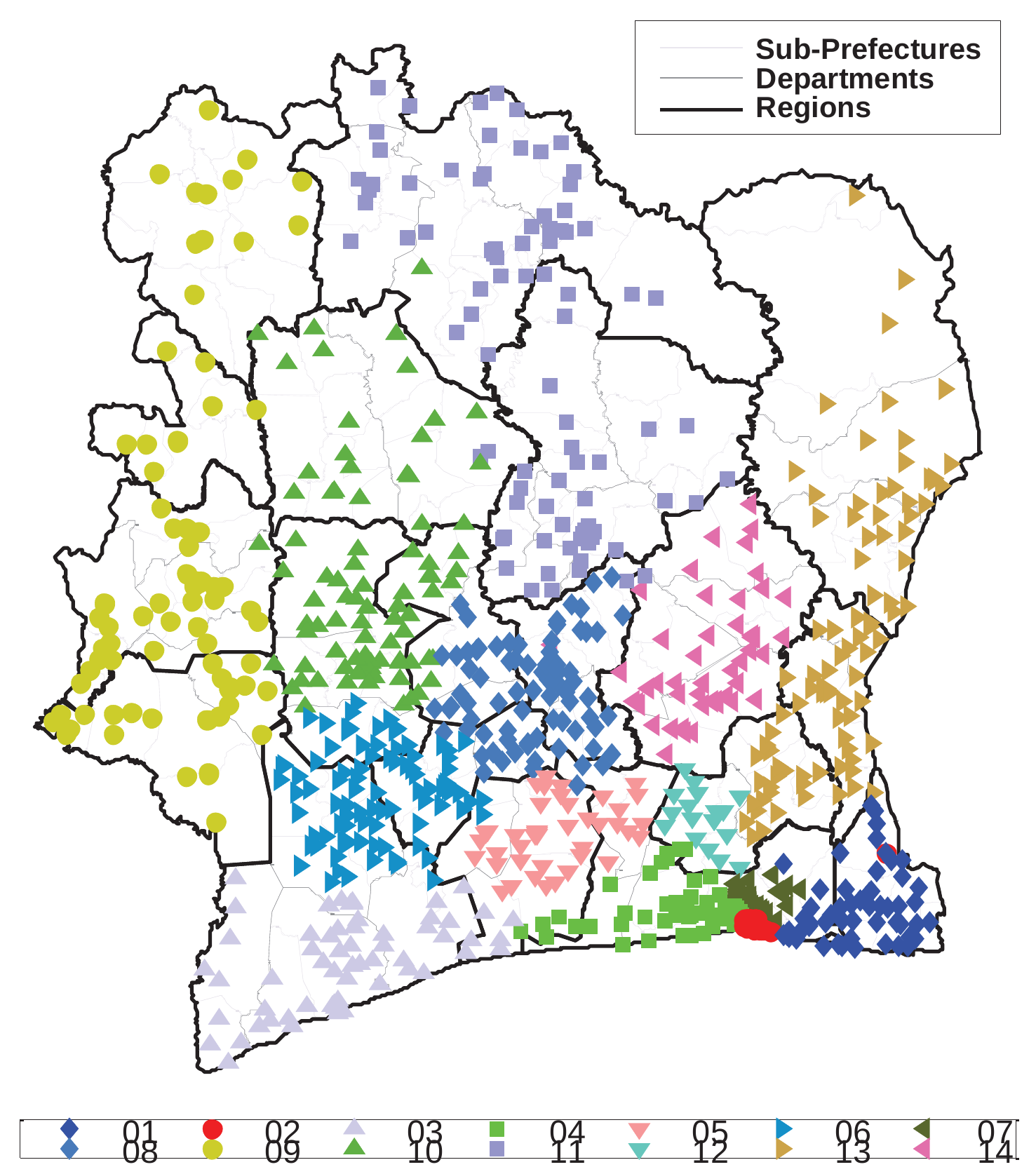}
}\qquad
\subfigure[][]{\label{fig7b}
\setlength{\abovecaptionskip}{0pt}%设置图形与标题之间的距离为0
\includegraphics[width=0.39\textwidth]{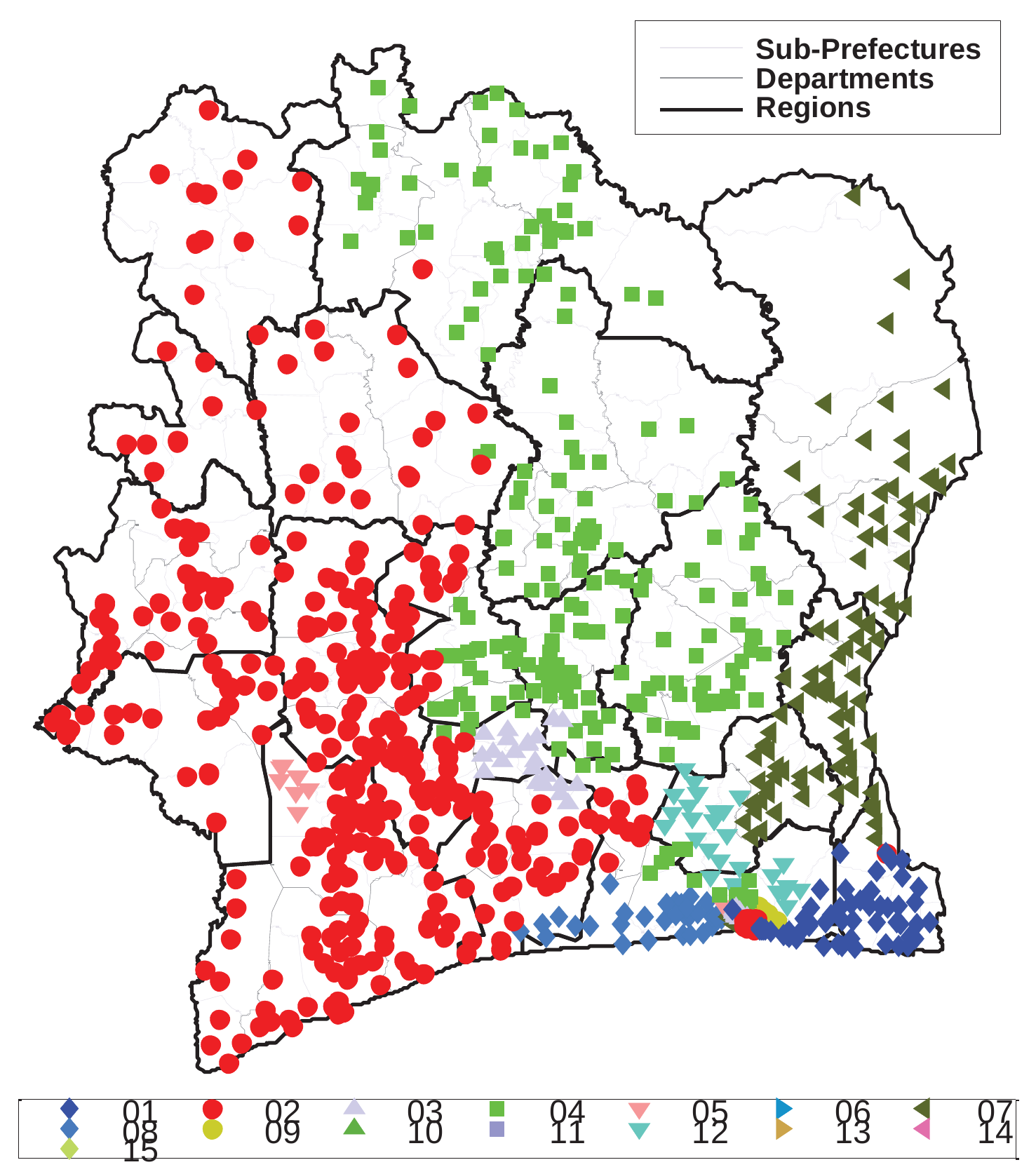}
}
\subfigure[][]{\label{fig7c}
\setlength{\abovecaptionskip}{0pt}%设置图形与标题之间的距离为0
\includegraphics[width=0.39\textwidth]{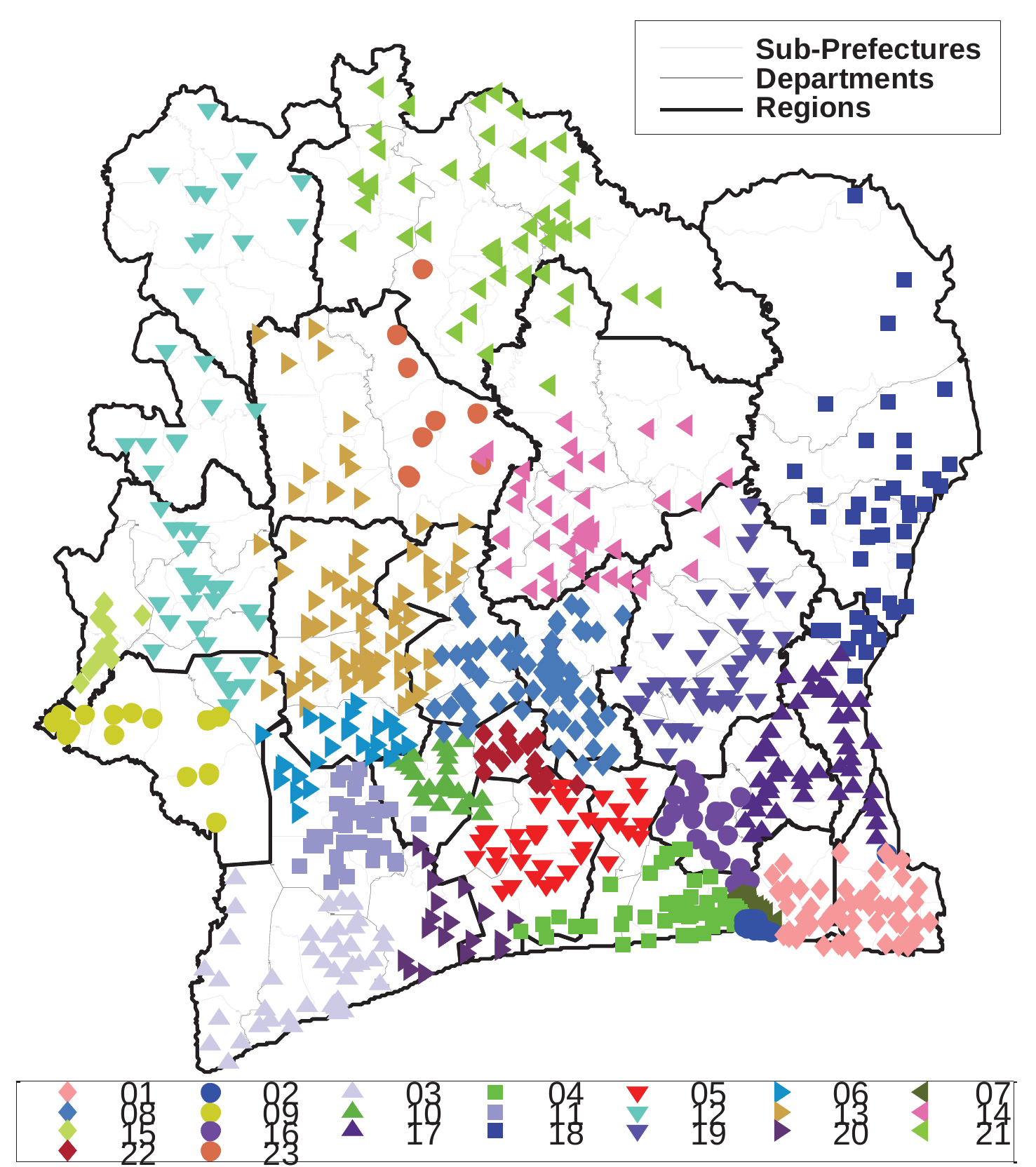}
}\qquad
\subfigure[][]{\label{fig7d}
\setlength{\abovecaptionskip}{0pt}%设置图形与标题之间的距离为0
\includegraphics[width=0.39\textwidth]{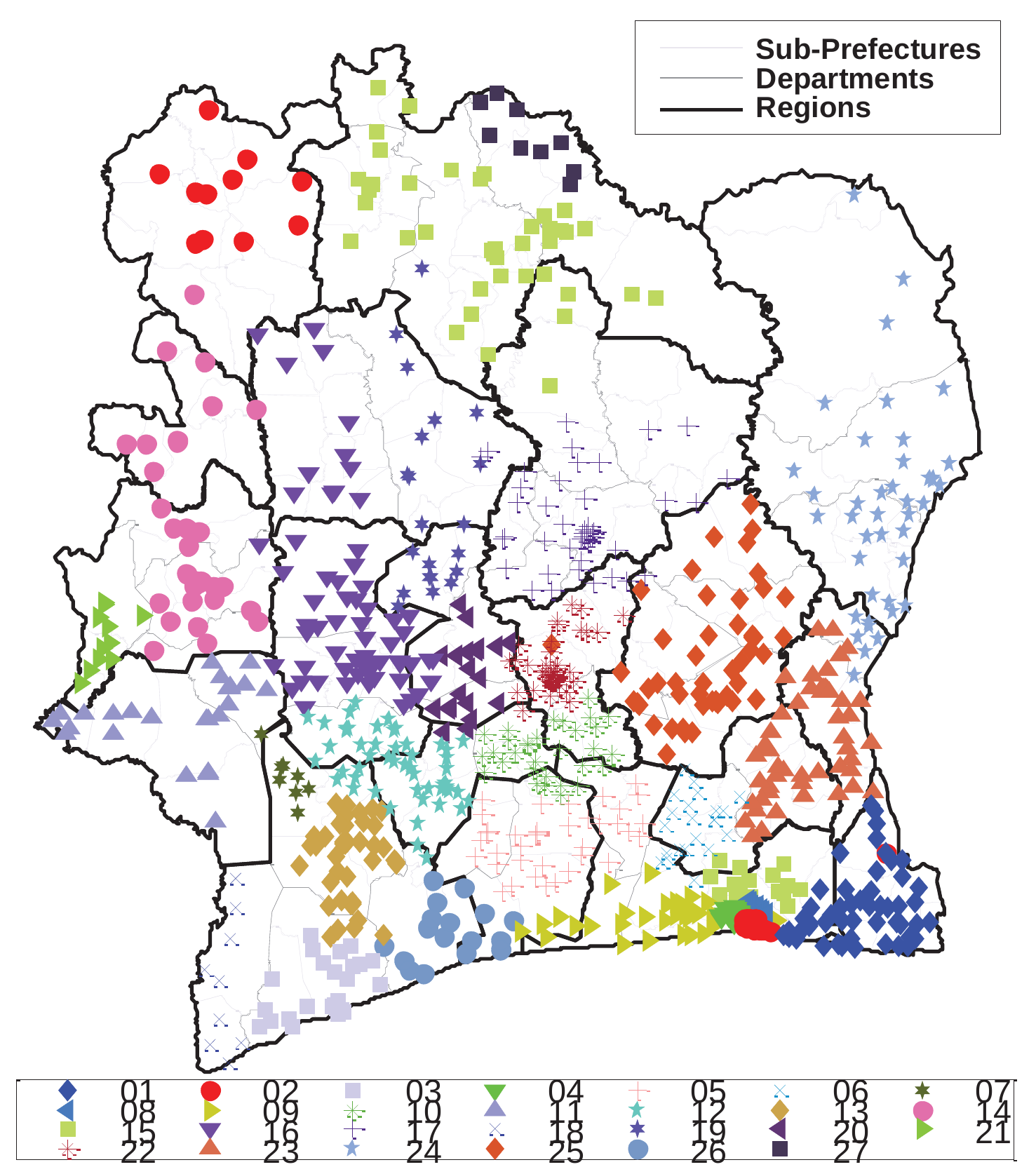}
}
\caption{\label{fig7} (Color online) The partitions of the D4D antenna network by (a) NG-Modularity, (b) Spa-Modularity, (c) Dist-Modularity at $\sigma=1.5\bar{d}$ (the consensus partition based on the alternative parameter selection), and (d) Dist-Modularity at $\sigma=0.7\bar{d}$.}
\end{figure*}
%\end{comment}

Fig.~\ref{fig5b} shows the assortativity effects of office location in the observed network, the Spa-Modularity null model, the NG-Modularity null model, and the Dist-Modularity null model. We can find that the Spa-Modularity null model assumes the same effect as what we observed in the network topology, while the Dist-Modularity null model assumes an even greater effect. Which is right? According to the Pearson's chi-squared test, the $p$-value for the null hypothesis that office location and practice are independent from each other is as high as 0.4848. Thus, under the pure effect of practice itself, links should not have a significant tendency to exist between nodes with the same office location. As a result, a reasonable explanation is that the actual effect of office location itself is greater than what we observed, and such effect is weakened due to the additional effect of practice. This can help us interpret why Spa-Modularity failed to detect the communities based on practice.

%$f(d)=\frac{1}{1+(d/\sigma)^2}$

\subsection{D4D Antenna Network}\label{sec4.3}
The third example is based on a dataset of anonymous records of cell phone calls between five million of Orange's customers in Cote d'Ivoire between Dec 1, 2011 and Apr 28, 2012 (Orange is the key brand of France Telecom, one of the world's leading telecommunications operators). This dataset was provided through the Data for Development (D4D) Challenge  \cite{D4DData}. From the dataset, we constructed an antenna-antenna network, which contains 1,216 nodes representing cell tower antennas of the country, and 689,909 links representing communications between antennas, with weight indicating the number of calls. Besides, we have coordinate information about geographical locations of antennas.

Many study showed that space has assortativity effect on network topology due to high cost associated to spatially distant links \cite{BarthelemySpatialNetworkReview}. Thus, we aim to detect communities beyond space by Dist-Modularity. The specific procedure is as follows. First, we used the great-circle distance to compute $d_{ij}$. Second, we specified $f(d)=(1+(d/\sigma)^2)^{-1}$, since this reciprocal function is reminiscent of the gravity models which have long been used to model space related interactions \cite{CarrothersGravityPotentialConcept,WilsonSpatialDistributionModel}. Note that this reciprocal function allows us to simulate the assortativity effect of space at different degrees by tuning $\sigma$ --- the assortativity effect in the null model gradually fades as $\sigma$ increases from $0$ to $+\infty$ (see Fig.~\ref{fig1}). Third, we conducted the alternative parameter selection for $\sigma \in [0.1\bar{d},2.0\bar{d}]$, with a step length of $0.1\bar{d}$, and finally arrived at the consensus partition for $\sigma=1.5\bar{d}$ (see Fig.~\ref{fig6a}).

%\begin{comment}
\begin{table}[!t]
\begin{center}
\begin{tabular}{|l|l|l|l|}
\hline
               & Level1 & Level2 & Level3 \\
\hline
NG-Modularity  & 0.7269 & 0.7133 & 0.6586 \\
Spa-Modularity  & 0.4096 & 0.3970 & 0.4196 \\
Dist-Modularity  & 0.7385 & 0.7556 & 0.7217 \\
\hline
\end{tabular}
\end{center}
\caption{The NMI scores between the partitions by the three Modularities and the partitions based on the three-level administrative divisions.}
\label{table2}
\end{table}
%\end{comment}

Fig.~\ref{fig7}(a)-(c) visualize the partitions by NG-Modularity, Spa-Modularity, and Dist-Modularity. We can find that the communities by Dist-Modularity and NG-Modularity are spatially compact. On the contrary, the communities by Spa-Modularity are spatially wide. For example, the community $\#02$ in Fig.~\ref{fig7b} covers almost the whole west part of the country. Fig.~\ref{fig6b} compares the space effect in the observed network, the Spa-Modularity null model, the NG-Modularity null model, and the Dist-Modularity null model. We can see that the Spa-Modularity null model assumes a space effect slightly different from what we observed in the network topology (the difference is due to the binning strategy for smoothing $p(d)$). On the other hand, the Dist-Modularity null model assumes that the space has a milder assortativity effect. This can explain why Dist-Modularity brings spatially more compact communities than Spa-Modularity.

%对于bb= 0 15 400 320
%第1个数字控制与左边的距离，越大与左边的距离越小
%第2个数字控制与下面的距离，越大与下边的距离越小
%第3个数字控制与右边的距离，越大与右边的距离越大，图片也越小
%第4个数字控制与上面的距离，越大与上面的距离越大
%\begin{comment}
\begin{figure}[!t]
\centering
\includegraphics[width=0.36\textwidth]{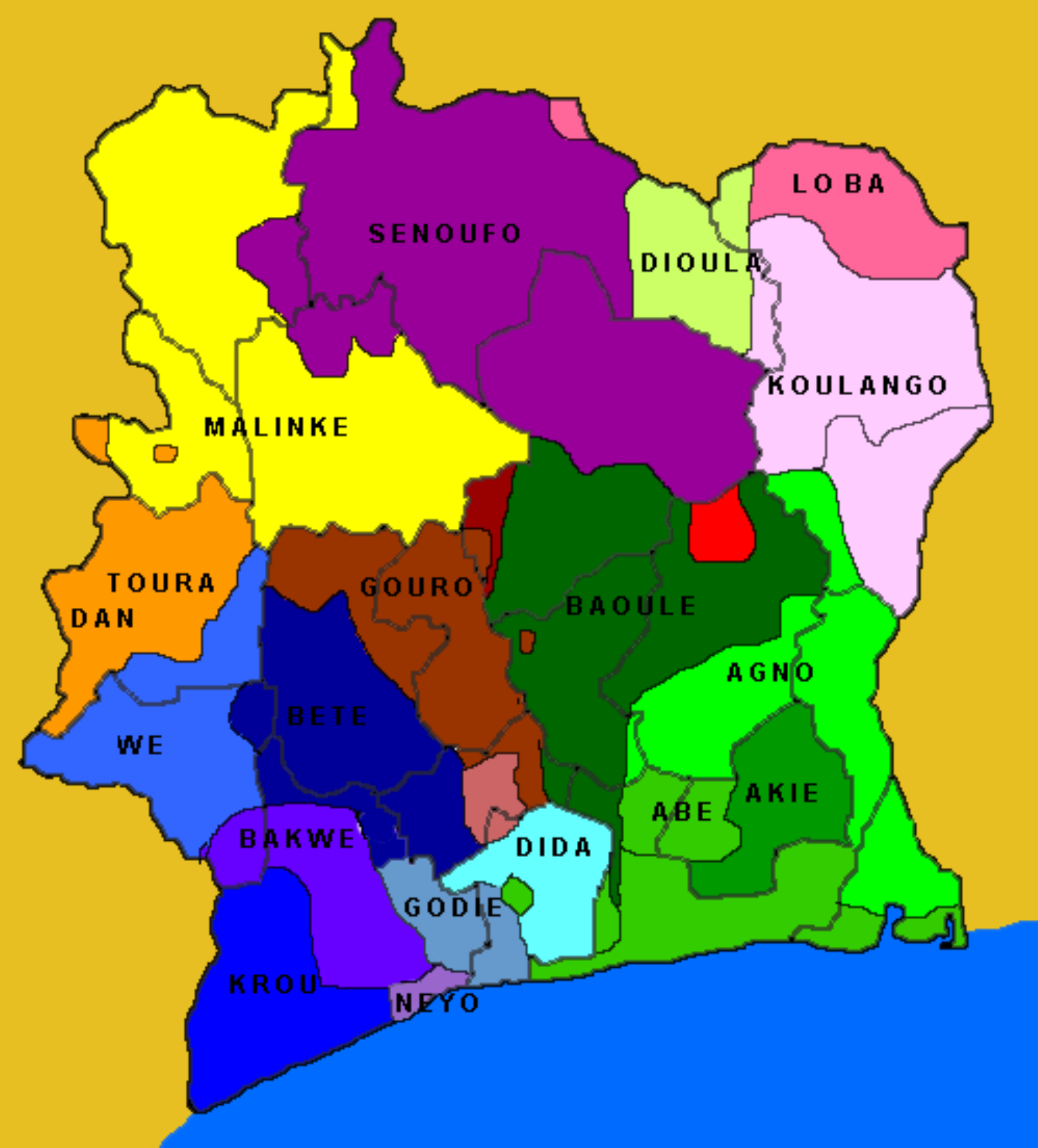}
\caption{\label{fig8} (Color online) The ethnic distribution in Cote d'Ivoire (image reprinted from http://fr.wikipedia.org/).}
\end{figure}
%\end{comment}
% http://fr.wikipedia.org/wiki/Crise\_politico-militaire\_en\_C$\mathrm{\hat{o}}$te\_d'Ivoire)

As an example, in Fig.~\ref{fig7d} we visualize the partition by Dist-Modularity at $\sigma=0.7\bar{d}$. A difference from the consensus partition at $\sigma=1.5\bar{d}$ is that some communities such as community $\#15$ are composed of several spatially distant groups of nodes which are themselves spatially compact. Note that the Dist-Modularity null model at $\sigma=0.7\bar{d}$ assumes a greater assortativity effect of space than that at $\sigma=1.5\bar{d}$. Thus, bringing together some spatially distant groups would contribute to a higher score of $\mathrm{Q}^{\mathrm{Dist}}$ at $\sigma=0.7\bar{d}$, and this results in the partition in Fig.~\ref{fig7d}. However, we can find that the key components of each community are spatially compact nodes. This indicates that the most critical factors that drive the network are highly correlated with space. Therefore, our partition based on the alternative parameter selection is reasonable.

According to 1998 Census, Cote d'Ivoire has 19 divisions at the region level, 50 at the department level, and 185 at the sub-prefecture level. The three-level administrative divisions are depicted by lines of different colors and width in Fig.~\ref{fig7}. It is interesting to find that the partition by Dist-Modularity coincides with the three-level divisions to a great extent. Indeed, as listed in Table~\ref{table2}, Dist-Modularity has the highest NMI scores between its partition and the three-level divisions. Note that the administrative divisions are highly correlated with the ethnic distribution of the country, as shown in Fig.~\ref{fig8}. Thus, the partition by Dist-Modularity is a good predictor of ethnic groups.

\section{Conclusion}\label{sec5}
In this paper, we focus on the problem of community detection beyond assortativity-related attributes $\rho$. A challenge of this problem is that we do not know to what extent the network topology is affected by $\rho$, and thus it is difficult to accurately simulate the effect of $\rho$ in the null model. We proposed Dist-Modularity which allows us to freely choose a function $f$ to simulate the effect of $\rho$. To apply Dist-Modularity to the problem, the key points are to probe $f$ using parameterized functions and conduct the alternative parameter selection to find a consensus partition. The success of our method lies in choosing the right form of function $f$ and giving a good estimation of the parameter interval. Thus, having a background knowledge about the network at study would be of help. We used three examples to demonstrate the effectiveness of our method.

Our method has significant practical applications. In particular, detecting terrorist communities beyond space may assist in tracking higher-level organizations, such as a logistics group that provides support to the terrorist cells. Shakarian (U.S. Military Academy) et al. are working with the agencies in the U.S. Department of Defense and developing a software based on Dist-Modularity for this emerging application \cite{ShakarianGeoDisperseCommunity}.

One issue of our method is the scalability. State of the art community detection and graph partitioning techniques which consider only network topology may scale to several hundred million nodes \cite{RaghavanLPA,DourisboureCommunityExtractionInLargeScaleWeb,UganderPartitioningMassiveGraph}. However, our method which considers both network topology and node attribute information requires $O(sn^2)$ time complexity. This limits applications to small and medium-sized networks. How to speed up the computation by high-performance computing resources, such as multi-cores, GPUs, clusters, is an important direction. This is left for our future work.

%\section*{Acknowledgments}
\begin{acknowledgments}
The authors are grateful to Prof. Alessandro Chessa (University of Cagliari) for providing the code of generating the synthetic spatial networks, to Prof. Tom A. B. Snijders (University of Oxford) for providing the dataset of partners' relationship in the New England law firm, and to France Telecom and Orange Cote d'Ivoire for providing the cell phone call dataset.
\end{acknowledgments}

\bibliography{MyRef31}% Produces the bibliography via BibTeX.

\end{document}